\documentclass[aps,prd,amsmath,amssymb,preprintnumbers,nofootinbib,a4paper,11pt]{revtex4}
\pdfoutput=1
\usepackage{amsthm}
\usepackage{graphicx}
	\graphicspath{{./NewFig/}}
\usepackage{url}
\usepackage[bookmarks, pagebackref=false]{hyperref}
\usepackage{color}
	\definecolor{rossoCP3}{cmyk}{0,.88,.77,.40}
		\definecolor{graa}{rgb}{0.8,0.8,0.8}
		\definecolor{blaa}{rgb}{0.2,0.2,0.6}
		\hypersetup{
			colorlinks, 
			bookmarksopen, 
			bookmarksnumbered,
			citecolor=blaa, 		
			linkcolor=rossoCP3,	
			urlcolor=rossoCP3,			
			}
\usepackage{dcolumn}
\usepackage{bm}
\usepackage{bbm}
\usepackage{pxfonts}

\usepackage{epsfig}
\usepackage{placeins}

\usepackage[ margin=5pt, font=small,labelfont=bf, justification=raggedright]{caption}
\usepackage{youngtab}
\usepackage{slashed}
\usepackage{subfig}

\newcommand{\beq}{\begin{eqnarray}}
\newcommand{\eeq}{\end{eqnarray}}

\newcommand{\bmp}{\noindent\begin{minipage}{16cm}}
\newcommand{\emp}{\end{minipage}\vskip 7mm} 

\def\lsim{\mathrel{\rlap{\lower4pt\hbox{\hskip1pt$\sim$}}
    \raise1pt\hbox{$<$}}}                
\def\gsim{\mathrel{\rlap{\lower4pt\hbox{\hskip1pt$\sim$}}
    \raise1pt\hbox{$>$}}}                

\begin{document}
\title{\Large  \color{rossoCP3}  The Physics of the $\theta$-angle \\for\\  Composite Extensions of the Standard Model} 
\author{Paolo Di Vecchia $^{\color{rossoCP3}{\clubsuit}}$}\email{divecchi@nordita.org} 
\author{Francesco Sannino$^{\color{rossoCP3}{\varheartsuit}}$}\email{sannino@cp3.dias.sdu.dk} 
\affiliation{$^{\color{rossoCP3}{\clubsuit}}$  \mbox{The Niels Bohr Institute, University of Copenhagen, Blegdamsvej 17, \\
DK-2100 Copenhagen, Denmark} \\ \mbox{Nordita, KTH Royal Institute of Technology and Stockholm University, Roslagstullsbacken 23, SE-106 91 Stockholm, Sweden
} \\ } 
 \affiliation{
$^{\color{rossoCP3}{\varheartsuit}}${ \color{rossoCP3}  \rm CP}$^{\color{rossoCP3} \bf 3}${\color{rossoCP3}\rm-Origins} \& the Danish Institute for Advanced Study {\color{rossoCP3} \rm DIAS},\\ 
University of Southern Denmark, Campusvej 55, DK-5230 Odense M, Denmark.
}

\begin{abstract}
We analyse the $\theta$-angle physics associated to extensions of the standard model of particle interactions featuring new strongly coupled sectors. We start by providing a pedagogical review of the $\theta$-angle physics for Quantum Chromodynamics (QCD) including also the axion properties. We then move to analyse composite extensions of the standard model elucidating the interplay between the new $\theta$-angle with the QCD one. We consider first QCD-like dynamics and then generalise it to consider several kinds of new strongly coupled gauge theories with fermions transforming according to different matter representations. Our analysis is of immediate use for different models of composite Higgs dynamics, composite dark matter and inflation.
\vskip .1cm
{\footnotesize  \it Preprint: CP$^3$-Origins-2013-34 \& DIAS-2013-34}
 \end{abstract}

\maketitle

     \newpage
     
     \tableofcontents
     
     \newpage
\section{Introduction} 
 
The Planck experiment \cite{Ade:2013zuv} has provided the most accurate determination to date of the composition of the universe. It has found that circa 95\% of the universe is made by unknown forms of matter and energy, while to describe
the remaining 5\% one needs at least three fundamental forces, i.e. Quantum Electrodynamics
(QED), weak interactions and Quantum Chromo Dynamics (QCD). Furthermore QCD, also known as strong interactions, is responsible for creating
the bulk of the bright mass, i.e. the 5\%. It is therefore natural to expect that to correctly describe
the rest of our universe, while providing a sensible link to the visible component, new forces will
soon emerge. There are at least three primary areas of research where new strong
dynamics can emerge. The first is the sector responsible for breaking spontaneously
the electroweak symmetry. The standard model Higgs sector in this scenario is expected to be replaced
by new strongly interacting dynamics. The second application is in the use of new strong dynamics to construct (near) stable dark matter candidates. Last but not the least  there is the intriguing possibility that even the mechanism behind inflation is powered by new strong dynamics. 

Not only QCD constitutes one of the pillars of the standard model of particle interactions, and accounts for the bulk of the visible matter in the universe, but it continues to pose formidable challenges both theoretically and phenomenologically. On the theoretical and experimental side we do not have yet a complete understanding of the strongly coupled infrared dynamics of the theory. 

Another puzzle is the experimental absence of otherwise theoretically legitimate $CP$ violating effects stemming from the topological sector of the theory known as the $\theta$-angle sector  \cite{Crewther:1979pi}. Topological sectors are known to be extremely relevant since they carry the underlying gauge theory imprint and can therefore help single out the underlying  dynamics
\cite{Witten:1979vv,Veneziano:1979ec,DiVecchia:1979bf,Crewther:1979pi,Rosenzweig:1979ay, DiVecchia:1980ve,Witten:1980sp,DiVecchia:1980gi,Narison:2008jp}. 

The purpose of this work is to provide a pedagogical review of the basic theoretical and phenomenological analyses of the $\theta$-angle physics for QCD, extend the analysis to other relevant gauge theories and, last but not the least, study the interplay between the $\theta$ physics of different extensions of the standard model of particle interactions featuring new strongly coupled sectors. 

In Section \ref{intro} we provide a pedagogical review of the $\theta$-angle physics for Quantum Chromodynamics (QCD) including also the axion properties. We then move to analyse composite extensions of the standard model elucidating the interplay between the new $\theta$-angles with the QCD one in Section \ref{extensions}.  We will present examples of how the introduction of new strongly coupled dynamics can affect the ordinary QCD $\theta$-angle physics.  In this section we will generalise the $\theta$-angle physics to consider several kinds of new strongly coupled gauge theories with fermions transforming in arbitrary matter representations. Last but not the least we will generalise the theories to include the lightest scalar state of the theory relevant both for QCD \cite{Beringer:1900zz} or its extensions were it be used for interpreting the composite state as the recently observed Higgs \cite{Aad:2012tfa,Chatrchyan:2012ufa} or the inflaton field \cite{Guth:1980zm,Linde:1981mu}. We conclude in Section \ref{conclusion} and in the Appendix \ref{AppendixA} we summarise some of the salient phenomenological imprints of the QCD $\theta$ physics.

Our analysis is of immediate use for different models of composite Higgs dynamics \cite{Sannino:2004qp,Dietrich:2005jn,Dietrich:2005wk,Dietrich:2006cm,Gudnason:2006mk,Ryttov:2008xe,Sannino:2009aw,Frandsen:2009mi}, composite dark matter  \cite{Gudnason:2006ug,Gudnason:2006yj,Gudnason:2006mk,DelNobile:2011je} and inflation \cite{Channuie:2011rq,Bezrukov:2011mv,Channuie:2012bv,Channuie:2013lla}.

\section{ Setting the stage: The QCD $\theta$ angle review} 
\label{intro}
Any extension of the standard model featuring a new $SU(N)$ gauge group can feature also a topological term. The topological term is added to the standard Yang-Mills Langrangian as follows:
\begin{equation}
L = - \frac{1}{4} F^{a}_{\mu \nu} F^{a \mu \nu} - \theta q (x) \ ,
\label{topo}
\end{equation}
where $a= 1,\dots, N^2-1$ with $N$ the number of colors of the given $SU(N)$ gauge theory and $ q(x)$ is the topological charge density given by:
\begin{equation}
q(x) = \frac{g^2}{32 \pi^2} F_{\mu \nu}^{a} {\tilde{F}}^{a \mu \nu} \ , \qquad
{\tilde{F}}^{\mu \nu} = \frac{1}{2} \epsilon^{\mu \nu \rho
  \sigma} F_{\rho \sigma} \ . 
\label{q}
\end{equation}
The additional term violates $CP$. This is easily understood since the topological term leads  to an operator of the form $\mathbf{E}^a \cdot \mathbf {B}^a$ when re-written directly in terms of the electric and magnetic field. Being a topological term, i.e. mathematically a volume term since the Lorentz indices are contracted via the four-dimensional fully antisymmetric tensor, it does not affect the classical equations of motions. Its physical effects derive from the interplay of field theory and quantum mechanics. In addition this operator, being of dimension four in mass dimensions, is renormalizable and therefore there is no theoretical reason forbidding its presence at the Lagrangian level. 
%

 In  QCD  this term is known as $\theta$-term and the associated $CP$ violation as
strong $CP$-violation  to distinguish it from sources of $CP$ violation due to the
 electroweak sector of the standard model. Experiments, however, do not observe any violation of strong $CP$  setting the very stringent upper bound $\theta < 10^{-9}$ .  In fact, as we shall see, the bound is for a specific linear combination of the  QCD $\theta$ angle and the argument of the determinant of the quark masses.

\subsection{QCD - Low Energy Effective Lagrangian}

To elucidate the physics of the theta angle the most efficient way is to use
the low energy effective Lagrangian of QCD featuring directly the  
pseudoscalar mesons and baryon composite states.  The $U(1)$ anomaly can be made explicit at the effective Lagrangian level which also allows to readily compute the relevant 
hadronic processes. Although the effective Lagrangian cannot be
explicitly derived from the fundamental QCD Lagrangian as it is, instead, the case of the  
$CP^{N-1}$ model~\footnote{See for instance Ref.~\cite{DiVecchia:1980gi} and
  references therein.}, one can nonetheless constrain its form by imposing the effective theory to faithfully respect both the anomalous and non-anomalous underlying QCD symmetries. 
 
The QCD Lagrangian with $N_f$ massless quark flavours possesses, at the classical level, a $U_L (N_f ) \times U_R (N_f )$ chiral symmetry that spontaneously
breaks to the diagonal vectorial subgroup $U_V (N_f )$. The pseudoscalar bosons
are the massless Goldstone bosons corresponding to the spontaneous breaking of
the chiral symmetry. 
 In the real world,
however,  
the light quarks are not massless.  They have a mass which can be considered small with respect to the intrinsic infrared  QCD scale $\Lambda_{QCD}$.  At low
energy the pseudoscalar bosons  are described by the following chiral 
Lagrangian:
\begin{equation}
L = \frac{1}{2} {\rm Tr} \left[ \partial_{\mu} U \partial_{\mu} U^{\dagger} \right] +
\frac{F_{\pi} }{2 \sqrt{2}} {\rm Tr} \left[ M ( U + U^{\dagger} ) \right]  \ ,
\label{effe}
\end{equation}
where $U$ contains the fields of the pseudoscalar mesons, that are
composite states of a quark and an antiquark:
\begin{equation}
U_{ij} = -\frac{2 \sqrt{2}m_i}{\mu_{i}^{2} F_{\pi}}
{\overline{\Psi}}_{R;i} \cdot \Psi_{L;j}\ , \qquad  \Psi_{R,L} = \frac{1
\pm  \gamma_5 }{2} \Psi \ ,
\label{uij}
\end{equation}
with $F_{\pi} = 95$~MeV,  the pion decay constant and $i, j = 1,\ldots, N_f$ the flavour index.  
The central dot in the first equation indicates the contraction of the colour indices.  We assume  the mass matrices of both the quarks and mesons to be diagonal
and real:
\begin{equation}
m_{ij} = m_i \delta_{ij}\ ,\qquad M_{ij} = \mu_{i}^{2} \delta_{ij} \ .
\label{mM}
\end{equation}
They are related by the Gell-Mann, Oakes and Renner relation \cite{GellMann:1968rz}:
\begin{equation}
\mu_{i}^{2} F_{\pi}^{2} = - 2 m_i <{\overline{\Psi}}_{R;i}  \cdot \Psi_{L;i} > \ ,
\label{GMOR}
\end{equation}
implying that the ratio $\frac{m_i}{\mu_{i}^{2}}$ is independent of $i$ since, in the limit of small masses, both $F_{\pi}$ and  the vacuum expectation value are flavour independent.
Notice that Eq. (\ref{uij}) is a consequence of Eq. (\ref{GMOR}) and of
the following equation:
\begin{eqnarray}
\frac{U_{ij}}{< U_{ij} >} =  2 \frac{ {\overline{\Psi}}_{R;i} \cdot 
\Psi_{L;j}}{ <{\overline{\Psi}}_{R; i} \cdot \Psi_{L;j}>} \ .
\label{upsi3}
\end{eqnarray}
It  can be easily checked that the first term of the Lagrangian in
Eq.~(\ref{effe}) is invariant, as the QCD Lagrangian without the term
involving the masses of the quarks, under the chiral $U_L (N_f ) \times U_R (N_f
)$ group that acts on $U$ as follows:
\begin{equation}
U \rightarrow g_L U g_R^{\dagger} \ ; \qquad  U^{\dagger} \rightarrow g_R U^{\dagger}
g_L^{\dagger}\ ; \qquad g_L^{-1} = g_L^{\dagger}; \qquad g_R^{-1} = g_R^{\dagger}   \ ,
\label{chi} 
\end{equation}
while the mass term breaks explicitly this symmetry precisely as the
quark mass matrix does in QCD. $g_{L/R}$ is a generic element of the first $U_{L/R}(N_f)$. The chiral
symmetry is spontaneously 
broken by imposing that the meson field satisfies the
constraint:
\begin{equation}
U U^{\dagger} = \frac{ F_{\pi}^{2}}{2}
\label{uuda}
\end{equation}
that implies:
\begin{equation}
U (x) = \frac{F_{\pi}}{\sqrt{2}} e^{ i \sqrt{2} \frac{\Phi (x)}{F_{\pi}}}
\qquad {\rm with} \qquad \Phi (x) = \Pi^{a} T^{a} + \frac{S}{\sqrt{N_f}} \ ,
\label{uexp}
\end{equation}
where  $T^a$ are the generators of $SU(N_f )$ in the fundamental
representation normalised as
\begin{equation}
{\rm Tr} [ T^{a} T^{b} ] = \delta^{ab} \ .
\label{nor}
\end{equation}
In the case of a $U(3)$ flavour symmetry $\Pi^{a} (x)$ corresponds to
the fields of the octet of the pseudoscalar mesons, while $S$ is a
$SU(3)$ singlet. In this case we get:
\begin{equation}
\Pi^{a} T^{a} = \frac{1}{\sqrt{2}} \left(\begin{array}{ccc}
   \pi^{0}+ \eta_8 /\sqrt{3} & \sqrt{2} \pi^{+} & \sqrt{2} k^{+} \\ 
   \sqrt{2} \pi^{-} & - \pi^{0} + \eta_{8}/\sqrt{3} & \sqrt{2} k^{0} \\
   \sqrt{2} k^{-} & \sqrt{2} {\bar{k}}^{0} & - 2\eta_{8}/\sqrt{3}
       \end{array}\right) \ .
\label{octet}
\end{equation}
The Lagrangian in Eq.~(\ref{effe}) does not reproduce correctly, however,  the
effect of the $U(1)$ axial anomaly since, apart from the mass term, is invariant under the axial $U(1)$. It is possible to take care of the axial anomaly, at the effective Lagrangian level, by adding an effective term containing the
topological charge density:
\begin{equation}
L = \frac{1}{2} {\rm Tr} \left[ \partial_{\mu} U \partial_{\mu} U^{\dagger} \right] +
\frac{F_{\pi} }{2 \sqrt{2}} {\rm Tr} \left[ M ( U + U^{\dagger} ) \right] + 
\frac{i}{2} q (x) {\rm Tr} \left[\log\frac{U}{U^{\dagger}} \right] \ .
\label{u1}
\end{equation}
Having introduced the background field $q (x)$, of mass dimension four, one can show, when taking the large number of QCD colours $N$ limit, that it is sufficient to add to the previous Lagrangian only a quadratic term in $q(x)$ since higher powers of
$q$ are suppressed in this limit. We arrive at the following Lagrangian:
\begin{equation}
L = \frac{1}{2} {\rm Tr} \left[ \partial_{\mu} U \partial_{\mu} U^{\dagger} \right]+
\frac{F_{\pi} }{2 \sqrt{2}} {\rm Tr} \left[ M ( U + U^{\dagger} ) \right] + 
\frac{i}{2} q (x) {\rm Tr} \left[\log\frac{U}{U^{\dagger}} \right] +
\frac{q(x)^2}{a F_{\pi}^{2}} \ .
\label{u1b}
\end{equation} 
We are ready to introduce explicitly the $\theta$ angle and study the physical consequences following the original derivations and results \cite{Witten:1979vv,Veneziano:1979ec,DiVecchia:1979bf,Crewther:1979pi,Rosenzweig:1979ay, DiVecchia:1980ve,Witten:1980sp} also reviewed in \cite{DiVecchia:1980gi}.  

%

\subsection{Adding the $\theta$ angle}

The $\theta$ angle multiplies the topological charge density and therefore the Lagrangian in
Eq.~(\ref{u1b}) is augmented by one more term as follows:
\begin{equation}
L = \frac{1}{2} {\rm Tr} \left[ \partial_{\mu} U \partial_{\mu} U^{\dagger} \right] +
\frac{F_{\pi} }{2 \sqrt{2}} {\rm Tr} \left[ M ( U + U^{\dagger} ) \right] + 
\frac{i}{2} q (x) {\rm Tr} \left[\log \frac{U}{U^{\dagger} }\right] +
\frac{q(x)^2}{a F_{\pi}^{2}} - \theta q(x) \ .
\label{u1c}
\end{equation} 
Since $q(x)$ is a background field, introduced to correctly saturate the axial anomaly and to take into account the $\theta$ term, it can now be eliminated through its equation of motion:
\begin{equation}
q (x) = \frac{a F_{\pi}^{2}}{2} \left[\theta - 
\frac{i}{2}  {\rm Tr} \left(\log U - \log U^{\dagger} \right) \right] \ .
\label{eqmo}
\end{equation}
Substituting the expression for $q(x)$ back in the effective Lagrangian we arrive at:
\begin{equation}
L = \frac{1}{2} {\rm Tr} \left[ \partial_{\mu} U \partial_{\mu} U^{\dagger} \right] +
\frac{F_{\pi} }{2 \sqrt{2}} {\rm Tr} \left[ M ( U + U^{\dagger} ) \right] - 
\frac{a F_{\pi}^{2}}{4} \left[\theta - 
\frac{i}{2}  {\rm Tr} \left[\log U - \log U^{\dagger} \right] \right]^2 \ .
\label{effec}
\end{equation}
Since $U U^{\dagger}$ is proportional to the identity matrix and the mass
matrix is diagonal the vacuum expectation value of $U$ must be:
\begin{equation}
< U_{ij} > = {e}^{-i \phi_i} \delta_{ij} \frac{ F_{\pi}}{\sqrt{2}} \ ,
\label{vevu}
\end{equation}
where, as we shall show, the quantities $\phi_i$ are determined by minimising the
energy. It is convenient to introduce the matrix $V$ that has a vacuum
expectation value proportional to the identity matrix:
\begin{equation}
U_{ij} = {e}^{-i \phi_i}V_{ij} \ , \qquad  < V_{ij} > =
\frac{F_\pi}{\sqrt{2}}  \delta_{ij} \ ,
\label{uv}
\end{equation}
and rewrite Eq. (\ref{effec}) in terms of the field $V$. We get $(
M_{ij}   \equiv  \mu_{i}^{2} \cos \phi_i \delta_{ij})$:
\[
L = \frac{1}{2} {\rm Tr} \left[ \partial_{\mu} V \partial_{\mu} V^{\dagger}  \right] +  
\frac{a F_{\pi}^{2}}{16} \left[ 
 {\rm Tr} \left[\log V - \log V^{\dagger} \right] \right]^2 + 
\frac{F_{\pi} }{2 \sqrt{2}} {\rm Tr} \left[ M    \left( V + V^{\dagger} - 
  \frac{2 F_{\pi}}{\sqrt{2}} \right)  \right] +
\]
\[
+ \frac{F_{\pi}^{2}}{2} \sum_{i=1}^{N_f} \mu_{i}^{2} \cos \phi_i -
\frac{a F_{\pi}^{2}}{4} \left( \theta - \sum_{i=1}^{N_f} \phi_i \right)^2  -
i \frac{F_{\pi}}{2 \sqrt{2}}{\rm Tr} \left[ \mu_{i}^{2} \sin \phi_i (V- V^{\dagger}) \right]
 + 
\]
\begin{equation}
+ i \left( \theta - \sum_{i=1}^{N_f} \phi_i \right) \frac{a  F_{\pi}^{2}}{ 4 }
 {\rm Tr} ( \log V - \log V^{\dagger} )  \ .
\label{effed}
\end{equation}
The angles $\phi_i$ are determined by minimising the total energy, namely:
\begin{equation}
E = \frac{F_{\pi}^{2}}{2} \left[
\frac{a}{2} ( \theta - \sum_{i=1}^{N_f} \phi_i )^2  - 
\sum_{i=1}^{N_f} \mu_{i}^{2} \cos \phi_i  \right] \ .
\label{ene}
\end{equation}
The minimisation yields:
\begin{equation}
 \mu_{i}^{2} \sin \phi_i = a \left( \theta - \sum_{i=1}^{N_f} \phi_i
 \right)\ , \qquad i=1 \dots N_f \ .
\label{mini}
\end{equation}
These equations determine the angles $\phi_i$ as a function of $a$ and $\theta$.
Substituting Eqs. (\ref{mini}) in Eq. (\ref{effed}) we get:
\begin{eqnarray}
L & = & \frac{1}{2} {\rm Tr} \left[ \partial_{\mu} V \partial_{\mu} V^{\dagger} \right] +  
\frac{a F_{\pi}^{2}}{16} \left(
 {\rm Tr} \left[ (\log V - \log V^{\dagger} \right] \right)^2 + 
\frac{F_{\pi} }{2 \sqrt{2}} {\rm Tr} \left[ M (\theta)   \left( V + V^{\dagger} - 
  \frac{2 F_{\pi}}{\sqrt{2}} \right)  \right] + \nonumber \\ 
&+& i  \left( \theta - \sum_{i=1}^{N_f} \phi_i \right) \frac{a  F_{\pi} }{ 2 \sqrt{2} }
\left( \frac{F_{\pi}}{\sqrt{2}} {\rm Tr} \left[ \log V - \log V^{\dagger} \right] - {\rm Tr} \left[ V- V^{\dagger} \right]
\right) - E_0 \ ,
\label{effedb}
\end{eqnarray}
where $E_0$ is the energy at the minimum. Since the matrix $V$ satisfies
the equation $V V^{\dagger} = \frac{F_{\pi}^{2}}{2}$, we can write V as follows:
\begin{equation}
V (x) = \frac{F_{\pi}}{\sqrt{2}} e^{ i \sqrt{2} \Phi (x)/F_{\pi}} \ , \qquad
  \Phi (x) = \Pi^{a} T^{a} + \frac{S}{\sqrt{N_f}} \ .
\label{uexpb}
\end{equation}
Substituting the above expressions in Eq.~(\ref{effedb})  we get:
\begin{eqnarray}
L &=& \frac{1}{2} {\rm Tr} \left[ \partial_{\mu} V \partial_{\mu} V^{\dagger} \right] -
\frac{a N_f}{2} S^2 + \frac{F_{\pi}^{2}}{2} {\rm Tr} \left[M(\theta) \left(\cos
    \frac{\sqrt{2} \Phi }{F_{\pi}} -1 \right) \right] + \nonumber \\
&+& \frac{a
  F_{\pi}}{\sqrt{2}} \left( \theta - \sum_{i=1}^{N_f} \phi_i \right) {\rm Tr}
\left[\frac{F_{\pi}}{\sqrt{2}} \sin \frac{\sqrt{2} \Phi}{F_{\pi}} -
  \Phi \right] - E_0 \ ,
\label{vexp}
\end{eqnarray}
where $\Phi$ is given in Eq. (\ref{uexpb}) and $M_{ij} (\theta) \equiv \mu_{i}^{2} \cos \phi_i
\delta_{ij} $.

The way to proceed is the following. First we have to solve Eq.s~(\ref{mini}) that determine $\phi_i$ as a function of $\theta, a$ and
$\mu_{i}^{2}$. Then insert them in the effective Lagrangian of
Eq.~(\ref{vexp}) that will depend on $\theta, a$ and
$\mu_{i}^{2}$. Before we proceed it is useful to show that the
quantities that we will extract from the previous effective Lagrangian
will be invariant under the shift  $\theta \rightarrow \theta + 2
\pi$.
This follows from the fact that, if we have found a solution $\phi_i
(\theta)$ of Eq.s~(\ref{mini}) then it is easy to show that also the
following will be a solution:
\begin{eqnarray}
\phi_{1} ( \theta + 2 \pi ) = \phi_1 (\theta) + 2 \pi \ , \qquad
\phi_i ( \theta + 2 \pi ) = \phi_i (\theta )\ , \qquad i=2 \dots N_f 
\label{nes}
\end{eqnarray}
But the physical quantities depend only on ${e}^{i \phi_i}$ and
therefore are invariant under a shift of $2 \pi$ of the $\theta$
angle.

It is also clear that strong $CP$ is conserved if 
$   \theta - \sum_{i=1}^{N_f} \phi_i =0$. This happens when:
\begin{enumerate}

\item{ $\theta =0 $ that implies that $\phi_i =0$,}

\item{ the mass of a quark flavour is zero}

\item{and $\theta = \pi$ for particular relations among the quark masses (see appendix). }

\end{enumerate}

\subsection{The Witten-Veneziano relation}

In order to get the Witten-Veneziano relation we have to consider the
theory without fermions. In this case the original effective
Lagrangian in Eq. (\ref{u1c}) becomes:
\begin{eqnarray}
L^{no ferm.} = \frac{q^2}{a F_{\pi}^{2}} - \theta q - i q J \ ,
\label{nofe}
\end{eqnarray}
where we have added an external source that is coupled to the
topological charge density $q$. From the previous expression one can
compute the partition function:
\begin{eqnarray}
Z ( J, \theta ) \equiv {e}^{-i W (J, \theta )} ={e}^{-i V_4 a
  F_{\pi}^{2} ( \theta + i J)^2/4} \ .
\label{zeta}
\end{eqnarray}
The vacuum energy is equal to:
\begin{eqnarray}
E (\theta) \equiv \frac{W(0, \theta)}{V_4} = \frac{ a F_{\pi}^{2}}{4}
\theta^2 
\label{vacene}
\end{eqnarray}
{From it} we get:
\begin{eqnarray}
\frac{d^2 E (\theta )}{d \theta^2}|_{\theta =0} = \frac{ a F_{\pi}^{2}}{2} \ .
\label{rel93}
\end{eqnarray}
On the other hand, neglecting the term with $M(\theta)$ in Eq.~(\ref{vexp}),  the mass of the singlet field can be obtained from
the effective Lagrangian in Eq. (\ref{vexp}) and it is equal to:
\begin{eqnarray}
M_{S}^{2} = a N_f \ .
\label{masi}
\end{eqnarray}
Putting together Eq.s (\ref{rel93}) and (\ref{masi}) we get the
Witten-Veneziano relation:
\begin{eqnarray}
M_{S}^{2} = \frac{2 N_f}{F_{\pi}^{ 2}}
\frac{d^2 E ( \theta )}{d \theta^2}|_{\theta =0} \ .
\label{wvrel}
\end{eqnarray}

\subsection{The  QCD axion}

{}From the analysis reviewed in the Appendix \ref{AppendixA},  we see that, if none of the  quark masses is exactly zero,  
  the $\theta$ angle must be very small.  If instead one of the quark masses were zero, $CP$ violation would be 
absent thanks to an exact classical symmetry (the chiral rotation of
the massless quark) which  allows to rotate $\theta$ away.  The latter solution is, however, disfavoured by lattice  and experimental low energy data \cite{Beringer:1900zz}. The strong $CP$ problem can, therefore, be stated in the following way: Within the standard model there is no natural explanation of why a parameter, unprotected by any symmetry, must vanish or being tuned to be very tiny. 

The solution to the strong $CP$ problem requires, therefore, to extend the standard model. {}For example, the Peccei-Quinn (PQ)  \cite{Peccei:1977hh,Peccei:1977ur} solution of the strong $CP$ problem includes new matter degrees of freedom. The essential property of the PQ model is that such an extension should  provide a new classically exact but quantum mechanically anomalous and spontaneously broken, $U(1)_{PQ}$ symmetry.

The low-energy effective action of such a theory will have to contain, besides the usual QCD degrees of freedom, an extra would-be Goldstone boson  related to the spontaneously broken $U(1)_{PQ}$ symmetry.
If we denote by $a_{PQ}$ the coefficient of the $U(1)_{PQ}$ anomaly
and  by $F_{\alpha}$ the scale of its spontaneous breaking (the analog 
of $F_{\pi}$), we can write down an effective action that 
incorporates all the relevant (anomalous and non-anomalous) Ward
identities. It is sufficient, indeed, to add a few terms to 
the effective Lagrangian of Eq. (\ref{u1c}) yielding \footnote{This analysis was performed in an unpublished paper by one of us (PDV) with G. Veneziano.
}:
\begin{eqnarray}
L &=& \frac{1}{2} {\rm Tr} \left[ \partial_{\mu} U \partial_{\mu} U^{\dagger} \right] + 
\frac{1}{2}  \partial_{\mu} N \partial_{\mu} N^{\dagger}  +
\frac{F_{\pi} }{2 \sqrt{2}} {\rm Tr} \left[ M ( U + U^{\dagger} ) \right]
+ \frac{q^2}{a F_{\pi}^{2}} - \theta q + \nonumber \\
&+& 
\frac{i}{2} q (x) \left({\rm Tr} \left[ \log U - \log U^{\dagger}\right]  + a_{PQ}(\log N - \log
  N^{\dagger} ) \right)  \ ,
\label{u1f}
\end{eqnarray}
where $U$ is given in \eqref{uexp} and 
\begin{eqnarray}
N (x) = \frac{F_{\alpha}}{\sqrt{2}} e^{ i \sqrt{2} \alpha
  (x)/F_{\alpha}} \ .
\label{enne}
\end{eqnarray}
Notice that, following our assumptions,  the only term that breaks 
 $U(1)_{PQ}$ is the one related to the anomaly.

Under the axial $U(1)$ and the additional $U(1)_{PQ}$ defined by:
\begin{eqnarray}
U \rightarrow {e}^{i \beta} U\ ; \qquad N \rightarrow {e}^{i \gamma} N \, ,
\label{tra45}
\end{eqnarray}
 the effective Lagrangian transforms as follows:
\begin{eqnarray}
\delta L = - \left(N_f \beta + a_{PQ} \gamma \right) q (x)  \ .
\label{tra46}
\end{eqnarray}
The Lagrangian is invariant if we impose $N_f \beta + a_{PQ}  \gamma =0$. This is an
anomaly-free $U(1)$ subgroup, whose spontaneous and explicit 
breaking (by quark masses) implies a new, pseudo-Goldstone boson, the 
(Peccei-Quinn-Weinberg-Wilczek) axion.

Proceeding as in the previous sections ($ < U_{ij} > = {e}^{- i
  \phi_i} \delta_{ij} F_{\pi}/\sqrt{2}$ and $<N> = {e}^{-i \phi/a_{PQ}}
  F_{\alpha}/\sqrt{2}$ ), we have to minimise the energy
given by:
\begin{eqnarray}
E = \frac{F_{\pi}^{2}}{2} \left[
\frac{a}{2} ( \theta - \sum_{i=1}^{N_f}  \phi_i  - \phi )^2  - 
\sum_{i=1}^{N_f} \mu_{i}^{2} \cos \phi_i  \right] \ .
\label{enea}
\end{eqnarray}
This gives
\begin{eqnarray}
a  \left( \theta - \sum_{i=1}^{N_f}  \phi_i - \phi \right) =
\mu_{i}^{2} \sin \phi_i \ ; \qquad \theta - \phi - \sum_{i=1}^{N_f} \phi_i =0 \ .
\label{min73}
\end{eqnarray}
The conditions above imply $ \phi_i =0$ and $\theta - \phi =0$.
In this case there is no dependence on the $\theta$ angle and no $CP$
violation because $\theta - \phi - \sum_{i=1}^{N_f} \phi_i =0$ (in
analogy, 
again, with the case of a single massless quark).

The mass matrix involving the axion and the components of $\Phi$
belonging to the Cartan subalgebra of $U(N_f )$ $(\Phi_{ij} = v_{i}
\delta_{ij})$  is given by:
\begin{eqnarray}
- \frac{1}{2} \left[ \sum_{i=1}^{N_f} \mu_{i}^{2} v_{i}^{2} +
  {a} \left(\sum_{i=1}^{N_f} v_i + b \alpha \right)^2  \right] \ ,
\label{mama}
\end{eqnarray}
where $ b \equiv a_{PQ} {F_{\pi}}/F_{\alpha}$. The masses of the
neutral mesons and of the axion are given by setting to zero the 
determinant of the
following matrix:
\begin{eqnarray}
\left( \begin{array}{cccccc} 
b^2 a - \lambda & b a & ba & ba & \dots & ba\\
ba & \mu_{1}^{2} + a - \lambda & a & a & \dots &        a  \\
ba & a &  \mu_{2}^{2} + a - \lambda & a  & \dots & a \\
\dots & \dots & \dots & \dots & \dots & \dots \\
ba & a & a & a  & \dots &  \mu_{N_f}^{2} + a - \lambda \end{array} \right) \ .
\label{matri}
\end{eqnarray}
The determinant of the previous matrix is equivalent to the one of the following matrix: 
\begin{eqnarray} 
\left( \begin{array}{cccccc} 
b^2 a - \lambda & b a & ba & ba & \dots & ba\\
\frac{\lambda}{b} & \mu_{1}^{2}  - \lambda & 0& 0 & \dots &        0  \\
\frac{\lambda}{b}& 0 &  \mu_{2}^{2} - \lambda & 0  & \dots & 0 \\
\dots & \dots & \dots & \dots & \dots & \dots \\
\frac{\lambda}{b} & 0 & 0 & 0  & \dots &  \mu_{N_f}^{2}  - \lambda \end{array} \right) \ ,
\label{matri2}
\end{eqnarray}
obtained from the first matrix by subtracting the first row divided by $b$ 
from all of the remaining rows.  By developing the determinant along 
the first row one derives:
\begin{eqnarray}
\lambda \left[ \frac{1}{a}+  \sum_{i=1}^{N_f} \frac{1}{\mu_{i}^{2} -
    \lambda}  \right] = b^2 \ .
\label{det45}
\end{eqnarray}
By solving for $\lambda$ one can determine the mass spectrum and its associated eigenstates involving the original axion and the pseudoscalars of the theory.  So far the analysis is completely general and applicable also to other non QCD theories. However, since phenomenologically for QCD,  
 $b <<1$  the lowest eigenvalue can be determined in a straightforward manner and corresponds to the mass of the QCD axion
\begin{eqnarray}
m_{\alpha}^{2} = \frac{b^2}{ \frac{1}{a} + \sum_{i=1}^{N_f}
  \frac{1}{\mu_{i}^{2}}}
  \sim \frac{b^2}{\frac{1}{\mu_{1}^{2}} +
\frac{1}{\mu_{2}^{2}}  } = 2 m_{\pi}^{2} b^2 \cdot \frac{m_1 m_2}{(m_1
  + m_2 )^2} \ ,
\label{maaxi}
\end{eqnarray}
where in the second passage we used the knowledge that the lightest quarks are the up and down, and invoked the chiral limit. In the last passage we used Eq.~(\ref{GMOR}) with $m_i$ the mass of the light quarks. Experimental constraints require that $F_{\alpha} \geq10^9$ GeV corresponding to an axion mass $m_{\alpha}
< 0.01$~eV.

The leading consequences for the meson and baryon physics of QCD are summarised in Appendix \ref{AppendixA}.
  
\section{The $\theta$ physics of minimal composite extensions of the Standard Model}
\label{extensions}
Having reviewed the salient properties of the QCD $\theta$-angle physics and associated strong $CP$ problem, we are now equipped to start investigating generalisations of the standard model featuring new strong dynamics sectors and associated new $\theta$-angles paying attention to their interplay with the QCD one.  

\subsection{QCD - like minimal composite extensions }
We consider here the class of composite extensions of the standard model constituted by a novel QCD-like theory (QCD') which couples to QCD via the mass term operator. A time-honoured class of models of this kind are minimal Technicolor extensions \cite{Sannino:2009za} according to which the Higgs sector of the standard model is replaced by a more fundamental interaction. Here by minimal we mean that the new theory does not carry ordinary colour. We also observe that the neutral new baryon of the theory can also be naturally identified with a dark matter candidate \cite{Nussinov:1985xr,Barr:1990ca,Gudnason:2006ug,Gudnason:2006yj,Foadi:2008qv,Frandsen:2009mi,DelNobile:2011je,Nardi:2008ix,Foadi:2008qv}. Another interesting possibility is that the new QCD' could describe directly and solely the dark matter sector  \cite{Sannino:2008nv,Buckley:2012ky}, i.e. a dark QCD which would still feel the weak interactions. The first lattice simulations of theories containing composite dark matter have only recently appeared \cite{Lewis:2011zb,Hietanen:2012sz,Appelquist:2013ms,Hietanen:2013fya}.  

Here we work in the low energy effective regime for both QCD and QCD'.  In this regime the low energy effective Lagrangian for QCD, as reviewed above, is:
\begin{eqnarray}
L_{QCD} = \frac{1}{2} {\rm Tr} \left[ \partial_{\mu} V \partial^{\mu} V^{\dagger} \right] -
\frac{af_{\pi}^{2}}{2} \left[ \theta - \frac{i}{2} {\rm Tr} \left[ \log \frac{V}{V^{\dagger}} \right]\right]^2 \ ,
\label{QCD} 
\end{eqnarray}
where $V = f_{\pi} {\rm e}^{i \frac{ \Phi}{f_{\pi}}  }$ with  $f_{\pi} \equiv \frac{F_{\pi}}{\sqrt{2}}$ with $F_{\pi} \sim 93$~MeV. For the sake of simplicity, we consider the case with two flavors.
$\Phi$ can be written in terms of the Pauli matrices $\tau_i$ and the identity matrix
\begin{eqnarray}
\Phi = \frac{1}{\sqrt{2}} \left( S + \Pi_i \tau_i   \right) \ .
\label{Phi}
\end{eqnarray}
The four matrices $\tau_i$ are the three Pauli matrices and the identity matrix. They are  normalized  such that ${\rm Tr} \left[ \tau_i \tau_j \right] = 2 \delta_{ij}$.

Analogously, the low energy effective Lagrangian for QCD' is:
\begin{eqnarray}
L_{QCD'} = \frac{1}{2} {\rm Tr} \left[ \partial_{\mu} U \partial^{\mu} U^{\dagger} \right] -
\frac{a' {f_{\pi}'}^{2}}{2} \left[ \theta' - \frac{i}{2} {\rm Tr} \left[ \log \frac{U}{U^{\dagger}} \right]\right]^2 \ .
\label{QCDprime}
\end{eqnarray}
We assume that the two theories communicate by means of the generalised {\it mass} term: 
\begin{eqnarray}
L_{mass} = f_{\pi} f_{\pi}'  {\rm Tr} \left[ \lambda ( U^{\dagger} V + V^{\dagger} U) \right] \ .
\label{Lmass}
\end{eqnarray}
$\lambda$ is a two by two diagonal matrix that we take to be real. In Technicolor extensions of the standard model such a term emerges naturally as a four-fermion operator from new sectors responsible for giving a mass to the standard model fermions.  

The complete Lagrangian reads \cite{DiVecchia:1980xq}
\begin{eqnarray}
L = L_{QCD} + L_{QCD'} + L_{mass}  \ .
\label{ELLETO}
\end{eqnarray}
In order to study the vacuum of the theory and the $CP$ violating terms we write the fields V and U as follows:
\begin{eqnarray}
&&V = X V_0 f_{\pi}\ , \qquad (V_0)_{ij}  \equiv {\rm e}^{- i \phi_j}\delta_{ij} \ , \qquad
U = Y U_0 f_{\pi}' \ , \qquad U_0 \equiv {\rm e}^{- i \phi_j '}\delta_{ij} \nonumber \\
&&X \equiv {\rm e}^{i \frac{\Phi}{f_{\pi}}  }\ , \qquad  \equiv {\rm e}^{i \frac{\Phi '}{f_{\pi}'}  } \ .
\label{V0U0} 
\end{eqnarray}
By inserting the previous expressions in Eq. (\ref{ELLETO}) we get:
\begin{eqnarray}
L & =  &\frac{f_{\pi}^{2} }{2} {\rm Tr} \left[ \partial_{\mu} X \partial^{\mu} X^{\dagger} \right] +
\frac{{ f_{\pi}'}^2 }{2} {\rm Tr} \left[ \partial_{\mu} Y \partial^{\mu} Y^{\dagger} \right]
- \frac{af_{\pi}^{2}}{2} \left[ \theta - \sum_j \phi_j - \frac{i}{2} {\rm Tr} \left[ \log \frac{X}{X^{\dagger}} \right]\right]^2 \nonumber \\
&- & \frac{a' {f_{\pi}'}^{2}}{2} \left[ \theta - \sum_j \phi_j '  - \frac{i}{2} {\rm Tr} 
\left[ \log \frac{Y}{Y^{\dagger}} \right]\right]^2 + f_{\pi}^{2}  {f_{\pi}'}^{2} {\rm Tr} \left[\Lambda Y^{\dagger} X + \Lambda^{\dagger} X^{\dagger} Y    \right] \ ,
\label{LL}
\end{eqnarray}
where
\begin{eqnarray}
\Lambda \equiv V_0 \lambda U_{0}^{\dagger} = {\rm e}^{-i ( \phi_i - \phi_i ') } \lambda_i \delta_{ij}
= \left(\cos ( \phi_i - \phi_i ') - i \sin ( \phi_i - \phi_i ') \right) \lambda_i \delta_{ij}  \ .
\label{LAM}
\end{eqnarray}
The angles $\phi_i$ and $\phi_i '$  are determined by minimizing the energy:
\begin{eqnarray}
E =  \frac{af_{\pi}^{2}}{2} \left( \theta - \sum_j \phi_j  \right)^2 +  \frac{a' {f_{\pi}'}^{2}}{2} \left( \theta' - \sum_j \phi_j '  \right)^2 -  2 f_{\pi}^{2}  {f_{\pi}'}^{2}\sum_j \lambda_j \cos ( \phi_j - \phi_j ')  \ .
\label{ENE}
\end{eqnarray}
We obtain the following equations:
\begin{eqnarray}
&- & af_{\pi}^{2} \left( \theta - \sum_j \phi_j  \right) +  2 f_{\pi}^{2}  f_{\pi}'^{2} 
\lambda_i \sin ( \phi_i - \phi_i ') =0\ , \qquad i=1,2  \nonumber \\
&- & a' f_{\pi}'^{2} \left( \theta' - \sum_j \phi_j '  \right) -  2 f_{\pi}^{2}  f_{\pi}'^{2} 
\lambda_i \sin ( \phi_i - \phi_i ') =0 \ , \qquad i=1,2  \ .
\label{mini1}
\end{eqnarray}
These equations lead to the following constraints:
 \begin{eqnarray}
&& af_{\pi}^{2}  \left( \theta - \sum_j \phi_j  \right) = - a' {f_{\pi}'}^{2}  \left( \theta' - \sum_j \phi_j '  \right)  \ ,  \qquad  \lambda_1 \sin( \phi_1 - \phi_1 ' ) = \lambda_2  \sin( \phi_2 - \phi_2 ' )  \ .
\label{mini2}
\end{eqnarray}
We can then write Eq. (\ref{LL}) as follows
\begin{eqnarray}
&&L = -E + \frac{f_{\pi}^{2} }{2} {\rm Tr} \left[ \partial_{\mu} X \partial^{\mu} X^{\dagger} \right]+
\frac{{ f_{\pi}'}^2 }{2} {\rm Tr} \left[ \partial_{\mu} Y \partial^{\mu} Y^{\dagger} \right]
+ \frac{af_{\pi}^{2}}{8} \left[   {\rm Tr} \left[ \log \frac{X}{X^{\dagger}} \right]\right]^2 
\nonumber \\
&&+ \frac{a' f_{\pi}'^{2}}{8} \left[    {\rm Tr} 
\left[\log \frac{Y}{Y^{\dagger}} \right]\right]^2 
+i \frac{af_{\pi}^{2}}{2} \left(\theta - \sum_j \phi_j \right)  
{\rm Tr} \left( \log \frac{X}{X^{\dagger}} \right)  \nonumber \\
&& + i \frac{a' f_{\pi}'^{2}}{2}  \left( \theta' - \sum_j \phi_j '  \right) \left[{\rm Tr} 
\left[ \log \frac{Y}{Y^{\dagger}} \right]\right] + f_{\pi}^{2}  f_{\pi}'^{2} 
{\rm Tr} \left[ M (\theta, \theta')  \left(Y^{\dagger} X +  X^{\dagger} Y -2 \right)    \right]
\nonumber \\
&&-i \frac{a f_{\pi}^{2}}{2} \left(\theta - \sum_j \phi_j \right) {\rm Tr} \left[  Y^{\dagger} X - X^{\dagger} Y \right] \ ,
\label{LL1}
\end{eqnarray}
where
\begin{eqnarray}
( M (\theta, \theta'))_{ij} = \lambda_i  \cos (\phi_i - \phi_{i}')   \delta_{ij} \ .
\label{EMME}
\end{eqnarray}
The previous Lagrangian can be written as the sum of a $CP$ conserving and a $CP$ violating term:
\begin{eqnarray}
L = L_{CPC} + L_{CPV}  \ ,
\label{CPCCPV} 
\end{eqnarray}
where (neglecting the constant term $-E_0$) 
\begin{eqnarray}
&& L_{CPC}  = \frac{f_{\pi}^{2} }{2} {\rm Tr} \left[ \partial_{\mu} X \partial^{\mu} X^{\dagger} \right] +
\frac{{ f_{\pi}'}^2 }{2} {\rm Tr} \left[ \partial_{\mu} Y \partial^{\mu} Y^{\dagger} \right]
+ \frac{af_{\pi}^{2}}{8} \left[   {\rm Tr} \left[ \log \frac{X}{X^{\dagger}} \right]\right]^2 
\nonumber \\
&& + \frac{a' f_{\pi}'^{2}}{8} \left[    {\rm Tr} 
\left[ \log \frac{Y}{Y^{\dagger}} \right]\right]^2   + f_{\pi}^{2}  f_{\pi}'^{2} 
{\rm Tr} \left[ M (\theta, \theta')  \left(Y^{\dagger} X +  X^{\dagger} Y -2 \right)     \right] \nonumber \\
&&   = \frac{f_{\pi}^{2} }{2} {\rm Tr} \left[ \partial_{\mu} X \partial^{\mu} X^{\dagger} \right] +
\frac{ f_{\pi}'^2 }{2} {\rm Tr} \left[ \partial_{\mu} Y \partial^{\mu} Y^{\dagger} \right]
  - a  S^2  - a'  {S'}^2
\nonumber \\
&& 
-   4 f_{\pi}^{2}  {f_{\pi}'}^{2} {\rm Tr} \left[ M (\theta, \theta') \sin^2 \left(\frac{\sqrt{f_{\pi}^{2} + {f_{\pi}'}^{2}} }{f_{\pi}  {f_{\pi}'} }  \frac{R}{2}  \right)\right]
\label{CPC}
\end{eqnarray}
and 
\begin{eqnarray}
&& L_{CPV} =  i \frac{a f_{\pi}^{2}}{2} \left( \theta - \sum_j \phi_j \right) {\rm Tr} \left[   \log \frac{X}{X^{\dagger}}  -\log \frac{Y}{Y^{\dagger}} -  \left(  Y^{\dagger} X - X^{\dagger} Y \right) \right]
\nonumber \\
&& =  a f_{\pi}^{2}  \left( \theta - \sum_j \phi_j \right) {\rm Tr} \left[ \frac{\sqrt{f_{\pi}^{2} + {f_{\pi}'}^{2}} }{f_{\pi}  {f_{\pi}'} } R - \sin \left( \frac{\sqrt{f_{\pi}^{2} + {f_{\pi}'}^{2}} }{f_{\pi}  f_{\pi}' } R\right) \right] \ .  
\label{CPV}
\end{eqnarray}
We have introduced the two following combinations
\begin{eqnarray}
&&R \equiv \frac{ f_{\pi} \Phi' - f_{\pi}' \Phi}{\sqrt{f_{\pi}^{2} + { f_{\pi}'}^2}}\ , \qquad 
T \equiv  \frac{ f_{\pi}' \Phi' + f_{\pi} \Phi}{\sqrt{f_{\pi}^{2} + { f_{\pi}'}^2}} \nonumber \\
&& \Phi' = \frac{  f_{\pi} R + f_{\pi}' T}{   \sqrt{ f_{\pi}^{2} + { f_{\pi}'}^2 } }\ , \qquad 
\Phi = \frac{  f_{\pi} T - f_{\pi}' R}{   \sqrt{ f_{\pi}^{2} + { f_{\pi}'}^2 } } \ .
\label{RT}
\end{eqnarray}
Notice that $L_{CPV}$  and also the mass term in the last line of Eq. (\ref{CPC}) depend 
only on $R$. The only dependence on $T$ appears in the kinetic terms and in the two mass 
terms of the flavour singlets $S$ and $S'$ in the next to the last line of Eq. (\ref{CPC}). 
This means that, independently from the form of the mass matrix,  the triplet of states contained in the matrix $T$ are always massless.  
One can introduce the electroweak gauge group in such a way\footnote{The standard model electroweak sector is introduced through the covariant derivatives:
\begin{eqnarray}
\partial_{\mu} X \Longrightarrow  D_{\mu} X = \partial_{\mu} X  + i g_2 A_{\mu} X - i g_1 X B_{\mu} \tau_3 \ , \qquad \partial_{\mu} Y \Longrightarrow  D_{\mu} Y = \partial_{\mu} Y  + i g_2 A_{\mu} Y - i g_1 Y B_{\mu} \tau_3
\label{covderiv}
\end{eqnarray}
and the addition of the gauge bosons kinetic terms:
\begin{eqnarray}
L_{gauge} = - \frac{1}{2} {\rm Tr} \left(   F_{\mu}  F^{\mu \nu } \right) - \frac{1}{4} B_{\mu \nu}B^{\mu \nu} 
+ \frac{f_{\pi}^{2}}{2} {\rm Tr } \left( D_{\mu} X D^{\mu} X^{\dagger} \right)  + \frac{ (f_{\pi} ')^{2}}{2} {\rm Tr } \left( D_{\mu} Y D^{\mu} Y^{\dagger} \right)  
\label{Lgaugekin}
\end{eqnarray}
where
\begin{eqnarray}
F_{\mu \nu} = \partial_{\mu} A_{\nu}  - \partial_{\nu} A_{\mu} + i g_2  [A_{\mu},  A_{\nu} ]\ , \qquad  B_{\mu \nu} = \partial_{\mu} B_{\nu} -  \partial_{\nu} B_{\mu} \ .
\label{FBmunu}
\end{eqnarray}
Here $A$ and $B$ are respectively the $SU(2)_L$ weak and $U(1)$ hypercharge gauge bosons.} 
that upon spontaneous symmetry breaking these three Goldstone bosons become, in the unitary gauge, the longitudinal degrees of freedom of  the gauge bosons 
$W^{\pm}$  and $Z$.  It is worth studying the mass of the pseudoscalar  mesons by concentrating only on the quadratic terms in the Lagrangian (\ref{CPC}).

The various fields are defined via:  
\begin{eqnarray}
&&R  = \frac{1}{\sqrt{2}} \left( R_a \tau^a + S_R \right) \ , \qquad T = \frac{1}{\sqrt{2}} \left( T_a \tau^a + S_T \right) \ , \nonumber \\ && \nonumber \\ 
&& R =\frac{1}{\sqrt{2}} \left(  \begin{array}{cc} R_3 + S_R  & R_1 - i R_2 \\
 R_1 + i R_2  & - R_3 +S_R \end{array} \right)   =     \left(  \begin{array}{cc} \frac{R_3 + S_R}{\sqrt{2}}  & R^- \\
 R^+  &  \frac{- R_3 +S_R}{\sqrt{2}}  \end{array} \right)     \ .
\label{phiphip}
\end{eqnarray}
the quadratic terms are given by
\begin{eqnarray}
L_2  &= &\frac{1}{2} {\rm Tr} \left(\partial_{\mu} T \partial^{\mu} T  \right) +
 \frac{1}{2} {\rm Tr} \left(\partial_{\mu} R \partial^{\mu} R  \right) -  a
  \frac{\left( f_{\pi} T_S - f_{\pi}' R_S\right)^2}{f_{\pi}^2 + { f_{\pi}'}^2}
 -  a'   \frac{\left(f_{\pi} R_S + f_{\pi}' T_S\right)^2}{ f_{\pi}^2 + { f_{\pi}'}^2}
- \left( f_{\pi}^2 + { f_{\pi}'}^2\right)   {\rm Tr} \left[ M (\theta, \theta')  R^2   \right]
 \nonumber \\
&=& \frac{1}{2} \sum_{a=1}^{3}  \left( \partial_{\mu} T_a \partial^{\mu} T_a \right) +
  \frac{1}{2} \partial_{\mu} T_S \partial^{\mu} T_S  +  \frac{1}{2} \partial_{\mu} R_S  \partial^{\mu} R_S + \frac{1}{2}  \sum_{a=1}^{3} \left( \partial_{\mu} R_a   \partial^{\mu} R_a   \right)   
  \nonumber \\
&& -    \frac{a f_{\pi}^{2} + a' { f_{\pi}'}^2}{f_{\pi}^2 + { f_{\pi}'}^2} T_{S}^{2}
 -   \frac{a { f_{\pi}'}^2 + a' f_{\pi}^{2}}{f_{\pi}^2 + { f_{\pi}'}^2} R_{S}^{2}   + 2 (a-a') 
 \frac{f_{\pi} f_{\pi}'}{f_{\pi}^2 + { f_{\pi}'}^2} T_S R_S  
 - \left( f_{\pi}^2 + { f_{\pi}'}^2\right)   {\rm Tr} \left[ M (\theta, \theta')  R^2   \right] \ ,
\label{L2}
\end{eqnarray}
where $R_S$ and $T_S$ are the $U(1)$ components of $R$ and $T$. 
If  we neglect the dependence on $\phi_i - \phi_i '$ in the mass matrix, the term with the mass is equal to:\
\begin{eqnarray}
&&- \left( f_{\pi}^2 + { f_{\pi}'}^2 \right) {\rm Tr } \left[ \left( \begin{array}{cc} \lambda_1 & 0 \\ 
0 & \lambda_2 \end{array} \right)  \left( \begin{array}{cc} \frac{R_3 + R_S}{\sqrt{2}} &  R_- \\ 
R_+ & \frac{R_S - R_3}{\sqrt{2}} \end{array} \right)  \left( \begin{array}{cc} \frac{R_3 + R_S}{\sqrt{2}} &  R_- \\ 
R_+ & \frac{R_S - R_3}{\sqrt{2}} \end{array} \right)    \right] \nonumber \\
&& = - \left( f_{\pi}^2 + { f_{\pi}'}^2 \right) {\rm Tr } \left[ \left( \begin{array}{cc} \lambda_1 & 0 \\ 
0 & \lambda_2 \end{array} \right)  \left( \begin{array}{cc} \frac{1}{2}(R_3 + R_S)^2 + R_- R_+ &  \sqrt{2} R_S R_- \\ 
\sqrt{2} R_S R_+ & \frac{1}{2}(R_S - R_3)^2 + R_- R_+ \end{array} \right)    \right] \nonumber \\
&& = - \left( f_{\pi}^2 + { f_{\pi}'}^2 \right)  \left[ \lambda_1  \left( \frac{1}{2}(R_3 + R_S)^2 + R_- R_+  \right)   + \lambda_2  \left( \frac{1}{2} (R_S - R_3)^2 + R_- R_+  \right)  \right]
\label{massterm6}
\end{eqnarray}
If $\lambda_1 = \lambda_2 \equiv \lambda$ then the terms that contribute to the mass are the following:
\begin{eqnarray}
&&-  \left( f_{\pi}^2 + {f_{\pi}'}^2 \right)  \lambda  \left[ R_{3}^{2} + R_{S}^{2} + 2 R_- R_+  \right]
 -    \frac{a f_{\pi}^{2} + a' {f_{\pi}'}^2}{f_{\pi}^2 + {f_{\pi}'}^2} T_{S}^{2}
 -   \frac{a {f_{\pi}'}^2 + a' f_{\pi}^{2}}{f_{\pi}^2 + {f_{\pi}'}^2} R_{S}^{2}   + 2 (a-a') 
 \frac{f_{\pi} f_{\pi}'}{f_{\pi}^2 + {f_{\pi}'}^2} T_S R_S
\nonumber \\ && 
\label{massma}
\end{eqnarray}
The triplet of states $R_a$ $(a=1,2,3)$ has mass squared equal to $2  \left( f_{\pi}^2 + { f_{\pi}'}^2 \right)  \lambda$, while the mass of the two singlet states is obtained by diagonalizing the following matrix:
\begin{eqnarray}
\left(  \begin{array}{cc}   \left( f_{\pi}^2 + { f_{\pi}'}^2 \right)   \lambda + 
 \frac{a { f_{\pi}'}^2 + a' f_{\pi}^{2}}{f_{\pi}^2 + { f_{\pi}'}^2} 
&   (a-a')   \frac{ f_{\pi} f_{\pi}'}{ f_{\pi}^2 + { f_{\pi}'}^2  } \\
  (a-a')   \frac{ f_{\pi} f_{\pi}'}{ f_{\pi}^2 + { f_{\pi}'}^2  }    & 
   \frac{a f_{\pi}^{2} + a' { f_{\pi}'}^2}{f_{\pi}^2 + { f_{\pi}'}^2}
  \end{array}  \right)
\label{massmavv}
\end{eqnarray}
As expected when $\lambda = 0$ the eigenvalues are respectively $a$ and $a'$ yielding the masses of the respective unmixed singlet pseudoscalars. 
%
%

If we add another explicit mass term, for example for the QCD' quarks, it is no longer possible to rotate away one linear combination of the theta angles and new sources of $CP$ violating operators will appear. This possibility is particularly interesting if the new QCD' physics is  used to give rise only to a dark sector.  

\subsection{ Quarks in arbitrary representations}
We now consider the case of  a QCD' theory in isolation - i.e. not yet coupled to the standard model or very weakly coupled -  with Dirac quarks transforming according to an arbitrary representation of the $SU(N)$ gauge group.  The $U(1)$ axial anomaly is given by:
\begin{eqnarray}
\partial_{\mu} J^{\mu}_{5} = 4N_f  c_R q(x)\ , \quad  {\rm with }\quad {\rm Tr} \left( \lambda^a \lambda^b \right) = c_R \delta^{ab}\ , \qquad q \equiv \frac{g^2}{32 \pi^2} F_{\mu \nu} {\tilde{F}}^{\mu \nu}  \ .
\label{axi}
\end{eqnarray}
For example, for the fundamental representation $c_R = \frac{1}{2}$ and for the two-index symmetric (antisymmetric) representations $c_R = \frac{N+2}{2}$
$\left(c_R = \frac{N-2}{2}\right)$. Explicitly for the two-index complex representations we have
 \begin{eqnarray}
\partial_{\mu} J^{\mu}_{5} = N_f \frac{N\pm 2}{2} \ \frac{g^2}{32\pi^2}  \epsilon^{\mu \nu \rho \sigma} F^a_{\mu \nu} F^a_{\rho \sigma} \equiv 2N_f \left( N \pm 2 \right) q \ .  \label{2-index}
\end{eqnarray}
One observes immediately that for the case of the antisymmetric representation, when  $N=3$ one recovers the fundamental representation. This is so because group theoretically the two-index antisymmetric representation for three colors is the fundamental representation \cite{Corrigan:1979xf,Kiritsis:1989ge,Armoni:2003gp,Sannino:2003xe,Sannino:2007yp}. 
For real representations such as the adjoint representation we have: 
 \begin{eqnarray}
\partial_{\mu} J^{\mu}_{5} = 2 N_w {N} \ \frac{g^2}{32\pi^2}  \epsilon^{\mu \nu \rho \sigma} F^a_{\mu \nu} F^a_{\rho \sigma} \equiv 2N_w  N q \ ,  \label{2-index}
\end{eqnarray}
with $N_w$ the number of Weyl fermions. Super Yang-Mills corresponds to $N_w=1$. The link at large $N$ between two-indices theories featuring one Dirac flavour and supersymmetric Yang-Mills was explored in \cite{Armoni:2003gp,Sannino:2003xe}. The application of higher dimensional representations for phenomenologically relevant candidates of new strong dynamics was put forward in \cite{Sannino:2004qp,Hong:2004td,Dietrich:2005jn,Dietrich:2006cm}. These theories are being investigated via first principle lattice simulations with interesting results \cite{Catterall:2007yx,Catterall:2008qk,Hietanen:2008mr,Catterall:2009sb,DelDebbio:2010hx,Kogut:2010cz,Kogut:2011bd,Bursa:2011ru,Giedt:2012rj} including the physical spectrum of the composite states \cite{Fodor:2012ty,Lewis:2011zb,Hietanen:2012sz, Lewis:2011zb,Hietanen:2012sz}. The phenomenology associated to minimal models of dynamical electroweak symmetry breaking is summarised in \cite{Appelquist:1998xf,Foadi:2007ue,Belyaev:2008yj,Andersen:2011yj,Franzosi:2012ih,Foadi:2012bb}. 
  
Since the pattern of chiral symmetry breaking for the case of two-index complex representations is identical to QCD, provided that the number of flavours is small enough that the underlying theory does not develop an infrared conformal fixed point \cite{Sannino:2004qp,Dietrich:2006cm,Ryttov:2007cx,Pica:2010mt,Pica:2010xq}, we can generalize Eq.~(\ref{u1c}) to take into account the associated anomaly in the following way:  
\begin{equation}
L = \frac{1}{2} {\rm Tr} \left[ \partial_{\mu} U \partial_{\mu} U^{\dagger} \right] +
\frac{F_{\pi} }{2 \sqrt{2}} {\rm Tr} \left[ M ( U + U^{\dagger} ) \right] + 
{i}  c_R q (x) {\rm Tr} \left[ \log U - \log U^{\dagger} \right] + \frac{q(x)^2}{aF^2_\pi} - \theta q(x)\ .
\label{u1later}
\end{equation}
For a given complex representation the pion decay constant scales at large $N$ as ${F_\pi}^2 \propto  d_R$ with $d_R$ the dimension of the representation which for the fundamental and two-index asymmetric/symmetric representations are respectively $N$ and $N(N\mp1)/2$. Technically $q(x)$ is an auxiliary field allowing to implement the axial transformations linearly. The introduction of the $\theta$ term is identical for any representation since appears in the Yang-Mills sector.  Eliminating the auxiliary field via its equation of motion the Lagrangian reads:
\begin{equation}
L = \frac{1}{2} {\rm Tr} \left[ \partial_{\mu} U \partial_{\mu} U^{\dagger} \right] +
\frac{F_{\pi} }{2 \sqrt{2}} {\rm Tr} \left[ M ( U + U^{\dagger} ) \right] - 
\frac{a F_{\pi}^{2}}{4} \left[\theta - 
{i}\,  c_R \,{\rm Tr} \left(\log U - \log U^{\dagger} \right) \right]^2 \ .
\label{effecr}
\end{equation}
Re-parametrizing the matrix $U$ with $V e^{-\phi}$, in order to minimise with respect to the abelian phases of $U$, we obtain:
\begin{eqnarray}
L & = & \frac{1}{2} {\rm Tr} \left[ \partial_{\mu} V \partial_{\mu} V^{\dagger} \right]+  
\frac{a F_{\pi}^{2}}{4} \, c_R^2\,\left(
 {\rm Tr} \left[\log V - \log V^{\dagger} \right] \right)^2 + 
\frac{F_{\pi} }{2 \sqrt{2}} {\rm Tr} \left[ M    \left( V + V^{\dagger} - 
  \frac{2 F_{\pi}}{\sqrt{2}} \right)  \right] \nonumber \\
&+& \frac{F_{\pi}^{2}}{2} \sum_{i=1}^{N_f} \mu_{i}^{2} \cos \phi_i -
\frac{a F_{\pi}^{2}}{4} \left( \theta -  2 c_R \sum_{i=1}^{N_f} \phi_i \right)^2  -
i \frac{F_{\pi}}{2 \sqrt{2}}{\rm Tr} \left[ \mu_{i}^{2} \sin \phi_i (V- V^{\dagger}) \right] \nonumber \\
&+& i c_R\left( \theta - 2 c_R \sum_{i=1}^{N_f} \phi_i \right) \frac{a  F_{\pi}^{2}}{ 2 }
 {\rm Tr} \left[ \log V - \log V^{\dagger} \right]  \ .
\label{effedcr}
\end{eqnarray}
This expression generalises (\ref{effed}) to a generic complex matter representation. Assuming that on the ground state  $\langle V \rangle = \langle V ^{\dagger} \rangle=  F_\pi/\sqrt{2}$ the total energy of the system is:
 \begin{equation}
E = \frac{F_{\pi}^{2}}{2} \left[
\frac{a}{2} ( \theta - 2 c_R \sum_{i=1}^{N_f} \phi_i )^2  - 
\sum_{i=1}^{N_f} \mu_{i}^{2} \cos \phi_i  \right] \ ,
\label{enecr}
\end{equation}
minimised for 
\begin{equation}
 \mu_{i}^{2} \sin \phi_i = 2 c_R \,  a  \left( \theta - 2c_R \sum_{i=1}^{N_f} \phi_i
 \right)\ ; \qquad i=1 \dots N_f \ .
\label{minicr}
\end{equation}

Substituting back in the Lagrangian we have:
\begin{eqnarray}
L & = & \frac{1}{2} {\rm Tr} \left[ \partial_{\mu} V \partial_{\mu} V^{\dagger} \right] +  
\frac{a F_{\pi}^{2}}{4} \, c_R^2 \left(
 {\rm Tr} \left[ \log \frac{V}{V^{\dagger}} \right] \right)^2 + 
\frac{F_{\pi} }{2 \sqrt{2}} {\rm Tr} \left[ M (\theta)   \left( V + V^{\dagger} - 
  \frac{2 F_{\pi}}{\sqrt{2}} \right)  \right] + \nonumber \\ 
&+& i\, 2c_R\, \left( \theta - 2c_R\, \sum_{i=1}^{N_f} \phi_i \right) \frac{a  F_{\pi} }{ 2 \sqrt{2} }
\left( \frac{F_{\pi}}{\sqrt{2}} {\rm Tr} \left[ \log\frac{ V}{V^\dagger}\right] - {\rm Tr} \left[ V- V^{\dagger} \right]
\right) - E_0 \ .
\label{effedb}
\end{eqnarray}
Using for $V$ equation (\ref{uexpb}) we obtain: 
\begin{eqnarray}
L &=& \frac{1}{2} {\rm Tr} \left[ \partial_{\mu} V \partial_{\mu} V^{\dagger} \right] -
{2 a N_f}\, c_R^2 S^2 + \frac{F_{\pi}^{2}}{2} {\rm Tr} \left[M(\theta) \left(\cos
    \frac{\sqrt{2} \Phi }{F_{\pi}} -1 \right) \right] + \nonumber \\
&+&  2c_R\frac{a
 F_{\pi}}{\sqrt{2}} \left( \theta - 2c_R\sum_{i=1}^{N_f} \phi_i \right) {\rm Tr}
\left[\frac{F_{\pi}}{\sqrt{2}} \sin \frac{\sqrt{2} \Phi}{F_{\pi}} -
  \Phi \right] - E_0. 
\label{vexpr}
\end{eqnarray}
From the previous action we deduce the mass of the pseudo scalar $S$: 
\begin{equation}
M^2_S= 4 a N_f c^2_R \ . 
\end{equation} 
We also have at large $N$  that $a \propto 1/d_R$, or equivalently  $a F^2_\pi$ is $N$ independent. This implies that at large $N$ the pseudoscalar $S$ becomes massless when fermions transform according to the fundamental representation while its mass becomes leading in $N$ for the two-index representations.

\subsection{Adding the lightest composite scalars}

It is, by now, well established that the correct description of the low energy $\pi\pi$ scattering data requires the introduction of the $\sigma$  state \cite{Sannino:1995ik,Harada:1995dc,Harada:1996wr,Black:1998wt} indicated as $f_0(500)$ by the particle data group \cite{Beringer:1900zz}. The latter makes use also of the dispersion relations results \cite{Colangelo:2001df,Ananthanarayan:2000ht,Kaminski:2002pe,Caprini:2005zr,GarciaMartin:2011cn,Moussallam:2011zg}   implementing the Roy equations \cite{Roy:1971tc} for $\pi\pi$ scattering. Historically this particle was introduced by Johnson and Teller \cite{Johnson:1955zz} and incorporated later in the Linear Sigma Model of Gell-Mann and Levy \cite{GellMann:1960np}. The Higgs sector of the standard model is  a Linear Sigma Model with the $\sigma$ state identified with the Higgs state. Within the standard model, however, the Higgs state is assumed to be elementary. 
%
Furthermore the Linear Sigma Model, is however, a specific realisation of the mechanism of spontaneous symmetric breaking which requires, for the standard model case, also the renormalizability of the model \footnote{Although the ATLAS and CMS collaborations have independently reported the discovery of a new particle \cite{Aad:2012tfa,Chatrchyan:2012ufa} with properties consistent with the standard model Higgs the burning question remains: Is the new particle state the standard model Higgs? It is tempting, by thinking fast, to accept the simplest paradigm, i.e. that it is the standard model Higgs.  After all, the standard model paradigm corresponds to the most minimal renormalisable model one can write able to break the electroweak symmetry preserving the $SU(2)_c$ custodial symmetry while giving masses to the standard model fermions, and it is compatible with the bulk of the experimental data \cite{Sannino:2013wla}. If the standard model paradigm is accepted then it becomes relevant to investigate its vacuum stability \cite{Degrassi:2012ry,Antipin:2013sga} making sure  that the quantum corrections do satisfy the Weyl consistency conditions determined in \cite{Antipin:2013sga,Antipin:2013pya}. According to these analyses the standard model is in a metastable state and can therefore tunnel to the true ground state located at much higher values of the Higgs field.  The stability of the potential, per se, is lost at around $10^{10}$~GeV reinforcing the idea that one needs to go beyond the standard model of particle interactions to have a more complete theory of nature.}.
  
However, the Linear Sigma Model, or any other effective Lagrangian, does not explain spontaneous symmetry breaking, at best parametrizes the phenomenon. Furthermore scalars are not fundamental representations of the Lorentz group, spin one-half fermions are. No elementary (pseudo)scalar has ever been discovered so far in Nature. It  would be the most important discovery made at the LHC. 
 
A composite Higgs and associate composite sector represent a natural solution to this problem. 
By composite, we mean composite by four-dimensional fermionic matter in the form of a strongly coupled gauge theory. One can, of course, enlarge the space of theories or the idea of compositeness, but should, at the same time, declare the standard model  problems is set to solve. 
 In technicolor, for example,  \cite{Weinberg:1979bn,Susskind:1978ms} the Higgs sector of the standard model is replaced by a new gauge dynamics featuring fermionic matter.  
 
 Because of the theoretical and phenomenological relevance of such a state both for QCD and the electroweak breaking sector of the standard model, as well as, any other extension of the standard model featuring composite dynamics, it is useful to extend the effective description investigated so far to incorporate this state. We refer to \cite{Belyaev:2013ida} for a recent relevant phenomenological analysis at the light of the LHC data. 
 
 Using as starting point the effective Lagrangian for any complex fermionic matter in a generic representation of the underlying $SU(N)$ gauge theory given in Eq.~\eqref{effecr} we extend it as follows:

\begin{eqnarray}
L_{\sigma}  & = & \frac{\kappa_D [\sigma]}{2} {\rm Tr} \left[ \partial_{\mu} U \partial_{\mu} U^{\dagger} \right] +
\frac{F_{\pi} }{2 \sqrt{2}} \, {\kappa_M [\sigma]} {\rm Tr} \left[ M ( U + U^{\dagger} ) \right] - 
\frac{a F_{\pi}^{2}}{4} \, \kappa_\theta [\sigma]\left[\theta - 
{i}\,  c_R \,{\rm Tr} \left(\log U - \log U^{\dagger} \right) \right]^2 + \nonumber \\ && \frac{1}{2} \partial_{\mu} \sigma \partial_{\mu} \sigma - \frac{m^2_\sigma}{2} \, \kappa_{m_\sigma}[\sigma] \, \sigma^2 \ .
\label{effecr-scalar}
\end{eqnarray}
with the $\kappa$  functions being Taylor expansions in $\sigma/(4\pi F_\pi)$ and the dimensionless coefficients of the expansion depend on the specific underlying gauge theory. We also have $\kappa[0]=1$ for any $\kappa$ function. There will also be higher derivatives in $\sigma$ but we consider only the leading order assuming that we are not too far, in the phenomenological processes, from the $\sigma$ mass production threshold.  The $\kappa_\theta$ term controls the theta physics of the scalar degree of freedom.

The generalisation to consider two coupled strongly interacting sectors can be achieved using as starting point, for example, the Lagrangian in Eq.~\eqref{ELLETO} with independent kappa functions for the two sectors, and therefore two independent scalar states, $\sigma$ and $\sigma^{\prime}$. The direct mixing between these two scalar states is induced by the generalised $L_{mass}$ term in the Lagrangian which now reads 
\begin{equation}
L_{mass} = \kappa_{mass}\left[\sigma, \sigma^{\prime}\right] f_\pi f^{\prime}_\pi {\rm Tr}\left[\lambda (U^{\dagger}V + V^{\dagger} U) \right] \ .
\end{equation}
The function $\kappa_{mass}$ depends on the specific extension coupling these two sectors and can be expanded simultaneously in $\sigma/(4\pi f_\pi)$ and $\sigma^{\prime}/(4\pi {f_\pi}^{\prime})$. 

\section{Conclusions}
\label{conclusion}
After having reviewed the $\theta$-angle physics, the associated strong $CP$ problem of QCD and its axion resolution, we considered extensions of the standard model featuring new strongly coupled sectors coupled to QCD.  In particular we elucidated the interplay between the new $\theta$-angle sector with the QCD one. Our analysis can be viewed as a stepping stone towards generic composite extensions of the standard model featuring new theta-angles. 

We have considered several kinds of new strongly coupled gauge theories with fermions transforming according to different matter representations of the underlying $SU(N)$ gauge theory. We have also shown how to generalise the framework to include the lightest scalar state of any strongly coupled theory (to be identified in QCD with the $\sigma$ state) and, for models of dynamical electroweak breaking, with the Higgs.
 
 Our analysis is of immediate use for different models of composite Higgs dynamics, composite dark matter and inflation.

\subsection*{Acknowledgements}
 The CP$^3$-Origins centre is partially funded by the Danish National Research Foundation, grant number DNRF90.

 \appendix

\section{Review of strong $CP$ violation phenomenological effects for QCD-like dynamics }

\label{AppendixA}
In this appendix we review, for completeness, how to obtain physically relevant observables for QCD induced by the presence of a nonzero $\theta$ angle.

\subsection{Strong $CP$ violating mesonic amplitudes}
We start  by minimising  Eq. (\ref{ene}) in the case of two flavours and in the limit where $ a
>> \mu_{1}^{2} ,  \mu_{2}^{2}$. In this case we must impose that
$\theta = \phi_1 + \phi_2$ and the minimisation equations become:
\begin{eqnarray}
\mu_{1}^{2} \sin \phi_1 = \mu_{2}^{2} \sin( \theta - \phi_1 ) \ .
\label{min94}
\end{eqnarray}
The solutions to the previous equation are:
\begin{eqnarray}
\sin \phi_1 = \frac{\mu_{2}^{2} \sin \theta}{\sqrt{\mu_{1}^{4} +
    \mu_{2}^{4} + 2\mu_{1}^{2} \mu_{2}^{2} \cos \theta }} \ , \qquad 
\sin \phi_2 = \frac{\mu_{1}^{2} \sin \theta}{\sqrt{\mu_{1}^{4} +
    \mu_{2}^{4} + 2\mu_{1}^{2}\mu_{2}^{2} \cos \theta}}  \ ,
\label{sin}
\end{eqnarray}
and
\begin{eqnarray}
\cos \phi_1 = \frac{\mu_{1}^{2} +\mu_{2}^{2} \cos \theta}{\sqrt{\mu_{1}^{4} +
    \mu_{2}^{4} + 2\mu_{1}^{2}\mu_{2}^{2} \cos \theta }}\ , \qquad
\cos \phi_2 = \frac{\mu_{2}^{2} + \mu_{1}^{2} \cos \theta}{\sqrt{\mu_{1}^{4} +
    \mu_{2}^{4} + 2\mu_{1}^{2} \mu_{2}^{2} \cos \theta }} \ .
\label{cos}
\end{eqnarray}
Computing the associated energy in Eq. (\ref{ene}) we get
\begin{eqnarray}
E (\theta) = - \frac{F_{\pi}^{2}}{2}\sqrt{\mu_{1}^{4} +
    \mu_{2}^{4} + 2\mu_{1}^{2} \mu_{2}^{2} \cos \theta } \ .
\label{ene56}
\end{eqnarray}
For equal masses $(\mu_{1} = \mu_2 = \mu)$  yields
\begin{eqnarray}
E (\theta) = - F_{\pi}^{2} \mu^2 \left| \cos \frac{\theta}{2} \right| \ .
\label{equama}
\end{eqnarray}
We find that both Eq.s (\ref{ene56}) and (\ref{equama}) are periodic of
period $ 2 \pi$ in $\theta$. Having solved the minimisation equation in the 
 $ a>> \mu_{1}^{2} ,  \mu_{2}^{2}$ limit we consider the first correction  
\begin{eqnarray}
\mu_{1}^{2} \sin \phi_1 = \mu_{2}^{2} \sin \phi_2 = a ( \theta -
\phi_1 - \phi_2 )
\label{eq93}
\end{eqnarray}
which can be determined by expanding around the large $a$ solution as follows
\begin{eqnarray}
\phi_{1,2} = {\bar{\phi}}_{1,2} + \epsilon \delta \phi_{1,2}\ , \qquad 
\epsilon = \frac{\mu_1 \mu_2}{a} \ .
\label{ansa}
\end{eqnarray}
One deduces
\begin{eqnarray}
\phi_1 = \bar{\phi}_1 - \epsilon \frac{\sin \theta}{R^3}
\frac{\mu_{2}^{2} +  \mu_{1}^{2} \cos \theta}{\mu_{1}^{2}} \ , \qquad 
\phi_2 = \bar{\phi}_2 - \epsilon \frac{\sin \theta}{R^3}
\frac{\mu_{1}^{2} +  \mu_{2}^{2} \cos \theta}{\mu_{2}^{2}} \ ,
\label{sol76}
\end{eqnarray}
where ${\bar{\phi}}_{1,2}$ is the large $a$ solution
\begin{eqnarray}
{\bar{\phi}}_{1} + {\bar{\phi}}_{2} = \theta \ , \qquad  R =
\sqrt{\frac{\mu_{1}^{4} +
    \mu_{2}^{4} + 2\mu_{1}^{2} \mu_{2}^{2} \cos \theta}{
\mu_{1}^{2}\mu_{2}^{2} }} \ .
\label{sol82}
\end{eqnarray}
Using the previous expression we can compute the 
$CP$ violating term contribution
\begin{eqnarray}
\theta - \phi_1 - \phi_2 = \epsilon \frac{\sin \theta}{R} =
\frac{\mu_{1}^{2}\mu_{2}^{2} \sin \theta}{ a \sqrt{\mu_{1}^{4} +
    \mu_{2}^{4} + 2\mu_{1}^{2} \mu_{2}^{2} \cos \theta }} \ .
\label{cpvio}
\end{eqnarray}
This contribution vanishes if $\theta =0$ or if $\mu_{1}^{2}$ and/or  $\mu_{2}^{2}$
are equal to zero. If $\mu_1 \neq \mu_2$ it is also zero for $\theta =
\pi$. But if $\mu_1 = \mu_2 \equiv \mu$ we get:
\begin{eqnarray}
\theta - \phi_1 - \phi_2 = \frac{\mu^2}{a} \sin \frac{\theta}{2}  = \frac{\mu^2}{a}  \ , \qquad {\rm for} \qquad \theta=\pi \ .
\label{pi}
\end{eqnarray}
One concludes that if $\mu_1 = \mu_2$ then $CP$ is violated also at $\theta =
\pi$.

{From} the $CP$ violating term in Eq. (\ref{vexp}) we can extract a
cubic term in the fields of the pseudoscalar mesons that is given by:
\begin{eqnarray}
- \frac{a \left( \theta - \sum_{i=1}^{N_f} \phi_i \right)}{3 \sqrt{2}
  F_{\pi}} {\rm Tr} \left[ \Phi^3 \right] \quad \longrightarrow  \quad - 
\frac{a \left( \theta - \sum_{i=1}^{N_f} \phi_i \right)}{\sqrt{3}  
  F_{\pi}} \pi^{+} \pi^{-} \eta_8 \ ,
\label{cpviob}
\end{eqnarray}
from which we extract the decay amplitude $ \eta_8 \rightarrow
\pi^{+} \pi^{-}$ given by
\begin{eqnarray}
T ( \eta \rightarrow \pi^{+} \pi^{-}) =  
\frac{a \left( \theta - \sum_{i=1}^{N_f} \phi_i \right)}{\sqrt{3} 
  F_{\pi}} = \frac{2 m_{\pi}^{2} (\theta )}{\sqrt{3} F_{\pi}} \cdot
\frac{\mu_{1}^{2}\mu_{2}^{2} \sin \theta }{\mu_{1}^{4} +
    \mu_{2}^{4} + 2\mu_{1}^{2} \mu_{2}^{2} \cos \theta} \ ,
\label{deca}
\end{eqnarray}
where
\begin{eqnarray}
 m_{\pi}^{2} (\theta ) = \frac{\mu_{1}^{2} \cos \phi_1 + \mu_{2}^{2} 
\cos \phi_2}{2} = \frac{1}{2} \sqrt{\mu_{1}^{4} +
    \mu_{2}^{4} + 2\mu_{1}^{2} \mu_{2}^{2} \cos \theta } \ .
\label{mpi}
\end{eqnarray}
For small values of $\theta$ we get
\begin{eqnarray}
T ( \eta \rightarrow \pi^{+} \pi^{-})  \sim  \frac{2
  m_{\pi}^{2}}{\sqrt{3} F_{\pi}} \frac{\theta}{\left(
  \sqrt{\frac{m_1}{m_2}} + \sqrt{\frac{m_2}{m_1}} \right)^2} \ ,
\label{smathe}
\end{eqnarray}
where $m_i$ is the quark mass related to the meson mass through
Eq. (\ref{GMOR}). Notice that in the previous calculation we have identified
$\eta_{8}$ with the particle state $\eta$~\footnote{The physical $\eta$ is the  linear combination $\eta = \cos \varphi  \, \eta_8 + \sin \varphi \, \eta_1$ of the $\eta_8$ and the isosinglet $\eta_1$ with a mixing angle $\varphi \sim 11$.}. 

{From} the previous equation we get:
\begin{eqnarray} 
\Gamma (  \eta \rightarrow \pi^{+} \pi^{-})  = \frac{\theta^2}{\left(
  \sqrt{\frac{m_1}{m_2}} + \sqrt{\frac{m_2}{m_1}} \right)^4 } \frac{m_{\pi}^4  \sqrt{m^2_{\eta} - 4m_{\pi}^2} }{12
 \pi F_{\pi}^2 m^2_{\eta}} \ .
\label{Gam}
\end{eqnarray}
Using $F_{\pi} = 95$~MeV, $m_{\pi} = 140 \, MeV$ and $m_{\eta}= 548$~MeV 
we get
\begin{eqnarray}
\Gamma (  \eta \rightarrow \pi^{+} \pi^{-}) ) = \frac{\theta^2}{ \left(
  \sqrt{\frac{m_1}{m_2}} + \sqrt{\frac{m_2}{m_1}} \right)^4}  \cdot 1.8 \, {\rm MeV}
  = \theta^2 \cdot 98.2
\,\,{\rm KeV}\ ,
\label{amp82}
\end{eqnarray}
and
\begin{eqnarray}
\frac{\Gamma (\eta \rightarrow \pi^{+} \pi^{-} ) }{\Gamma_{tot}} =
68 \,\,\theta^2 \ .
\label{ampli}
\end{eqnarray}
{From} experiments we have
\begin{eqnarray}
\frac{\Gamma (\eta \rightarrow \pi^{+} \pi^{-} ) }{\Gamma_{tot}} < 1.3
\cdot 10^{-5} \ ,
\label{exp}
\end{eqnarray}
that yields an upper limit on the value of $\theta < 4.4 \times 10^{-4}$. We will
get a much better limit from the electric dipole moment of the
neutron. The decay amplitude of $ \eta \rightarrow \pi^{+} \pi^{-}
$ is zero for $\theta=0$ and $\pi$ given that $ \mu_{1}^{2} \neq  \mu_{2}^{2}$. For extensions of the standard model where these masses are not yet determined we recall that if $ \mu_{1}^{2} =   \mu_{2}^{2}$ the corresponding process is not vanishing anymore at $\theta = \pi$.

In the previous analysis we have assumed that there are only two quark
flavours. In the case of three flavours one finds that 
\begin{enumerate} 
\item{If $| \mu_{2}^{2} - \mu_{1}^{2} |  \mu_{3}^{2} >   
\mu_{1}^{2} \mu_{2}^{2}$ then $CP$ is conserved at $\theta = \pi$}
\item{If $| \mu_{2}^{2} - \mu_{1}^{2} |  \mu_{3}^{2} <  \mu_{1}^{2} \mu_{2}^{2}$ then  $CP$ is violated at $\theta =\pi$. }
\end{enumerate}

{From} the meson mass matrix one can easily get the mass of the
pseudoscalar mesons as a function of the angle $\theta$. One gets:
\begin{eqnarray}
m_{\pi^{0} , \pi^{\pm}}^{2} = \frac{\mu_{1}^{2} \cos \phi_1 + \mu_{2}^{2} 
\cos \phi_2}{2}\ , \qquad m_{k^{\pm}}^{2} = 
\frac{\mu_{1}^{2} \cos \phi_1 + \mu_{3}^{2} 
\cos \phi_3}{2}  \ ,
\label{mass}
\end{eqnarray}
and 
\begin{eqnarray}
m_{k^{0}; {\bar{k}}^{0}}^{2} = 
\frac{\mu_{2}^{2} \cos \phi_2 + \mu_{3}^{2} 
\cos \phi_3}{2} \ .
\label{mass2}
\end{eqnarray}
These relations imply
\begin{eqnarray}
R (\theta)  \equiv \frac{m^{2}_{k^0} - m^{2}_{k^+} - m^{2}_{\pi^0} +
  m^{2}_{\pi^+} }{m_{\pi}^{2}} = \frac{\mu_{2}^{2} \cos \phi_2 -
  \mu_{1}^{2} \cos \phi_1}{\mu_{2}^{2} \cos \phi_2 +
  \mu_{1}^{2} \cos \phi_1 } = \frac{( \mu_{2}^{2} - \mu_{1}^{2})
( \mu_{2}^{2} + \mu_{1}^{2}) }{\mu_{1}^{4} +
    \mu_{2}^{4} + 2\mu_{1}^{2} \mu_{2}^{2} \cos \theta   } 
\label{mas67} 
\end{eqnarray}
where we have used Eq.s (\ref{cos}). In particular one deduces
\begin{eqnarray}
R (\theta =0 ) = \frac{\mu_{2}^{2} - \mu_{1}^{2} }{\mu_{2}^{2} +
  \mu_{1}^{2} }\ , \qquad R (\theta =\pi ) = \frac{\mu_{2}^{2} + \mu_{1}^{2}}{\mu_{2}^{2} -
  \mu_{1}^{2} } \ .
\label{rat9}
\end{eqnarray}
Experimentally $R \simeq 0.26$ which is consistent with $\theta =0$. The
ratio of masses for the two lightest quarks is determined from the
following relation
\begin{eqnarray}
\frac{m_1}{m_2} = \frac{\mu_{1}^{2}}{ \mu_{2}^{2}} = \frac{2
  m_{\pi^0}^{2} - m_{\pi^+}^{2} +  m_{k^+}^{2} -  m_{k^0}^{2} }{
  m_{k^0}^{2} - m_{k^+}^{2} +  m_{\pi^+}^{2}  } \simeq 0.56 , \quad {\rm for } \quad \theta=0 \ .
\label{rat8}
\end{eqnarray}
For the sake of completeness we provide also the ratio between the mass
of the strange and that of the down quarks:
\begin{eqnarray}
\frac{m_3}{m_2} = \frac{\mu_{3}^{2}}{ \mu_{2}^{2}} = \frac{
 m_{k^0}^{2}  - m_{\pi^+}^{2} +  m_{k^+}^{2}  }{
  m_{k^0}^{2} - m_{k^+}^{2} +  m_{\pi^+}^{2}  } \simeq 20.18
\label{rat7}
\end{eqnarray}

\subsection{Strong $CP$ violating  amplitudes with baryons}

In order to compute the $CP$ violating terms involving baryons  we add  to the effective Lagrangian terms involving
baryons. The baryons belong to an octet of $ SU_V(3)$  and are described
by the following matrix:
\begin{eqnarray}
B = \left(\begin{array}{ccc}
\frac{\Sigma^{0}}{\sqrt{2}}+ \frac{\Lambda}{\sqrt{6}} & \Sigma^{+} & p \\ 
\Sigma^{-} & -\frac{\Sigma^{0}}{\sqrt{2}}+ \frac{\Lambda}{\sqrt{6}} & n \\
\Xi^{-} & {{\Xi}}^{0} &  - 2 \frac{\Lambda}{\sqrt{6}}
       \end{array}\right) \ .
\label{ba}
\end{eqnarray}
Here $B$ is a Dirac spinor and, being a matter field, transforms naturally under the $SU(3)_V$ diagonal vector subgroup, 
\begin{equation}
B \rightarrow k B k^{\dagger} \ , \quad {\rm with} \quad k \in SU(3)_V \ .
\label{Btransform}
\end{equation}
The constraint equation linking $k$ to the underlying pion dynamics and the original $SU(3)\times SU(3)$ global symmetry is obtained imposing 
\begin{equation}
g_L \xi(\Phi) k^{\dagger}(\Phi,g_L, g_R) = k(\Phi,g_L, g_R) \xi(\Phi) g_R^{\dagger}  \ , \quad {\rm with} \quad  \xi \xi \equiv {U}  \frac{\sqrt{2}}{F_\pi} \ .
\end{equation} 
Under the chiral $SU_L(3)
\times SU_R(3)$ we can define purely left and right globally transforming baryon fields:
\begin{eqnarray}
R \equiv  \frac{1+\gamma_5}{2} \xi^{\dagger} B \xi \rightarrow g_R R g_R^{\dagger}\ , \qquad
L \equiv  \frac{1-\gamma_5}{2} \xi B  \xi^{\dagger}\rightarrow g_L L g_L^{\dagger} \ .
\label{rRL}
\end{eqnarray}
The meson fields transform as in Eq. (\ref{chi}) and therefore the, relevant to us, Lagrangian involving baryons can be written as follows
\begin{eqnarray}
L_{bar} = {\rm Tr} \left[{\bar{B}} i \gamma^{\mu} \partial_{\mu} B \right] -
\frac{\sqrt{2} \alpha}{F_{\pi}} {\rm Tr} \left[ {\bar{L}} U R U^{\dagger} + {\bar{R}}
  U^{\dagger} L U \right] 
+ \delta {\rm Tr} \left[ {\bar{L}} U R M
+ {\bar{R}} U^{\dagger} L M^{\dagger} \right] + 
\gamma {\rm Tr} \left[ {\bar{L}} M^{\dagger} R U^{\dagger} + {\bar{R}}
  M L U \right] \ .
\label{bar} \nonumber \\
\end{eqnarray}
In terms of $\xi$ and $B$ reads: 
\begin{eqnarray}
L_{bar} &=& {\rm Tr} \left[{\bar{B}} i \gamma^{\mu} \partial_{\mu} B \right] -
\alpha \frac{F_\pi }{\sqrt{2}} {\rm Tr} \left[ {\bar{B}} B \right] 
+  \delta \frac{F_\pi }{2 \sqrt{2}} {\rm Tr} \left[ {\bar{B}} B  (\xi M \xi + \xi^{\dagger}M^{\dagger} \xi^{\dagger}) \right]+  \delta \frac{F_\pi }{2 \sqrt{2}} {\rm Tr} \left[ {\bar{B}} \gamma_5 B  (\xi M \xi - \xi^{\dagger}M^{\dagger} \xi^{\dagger}) \right]  \nonumber \\
&+&   \gamma \frac{F_\pi }{2 \sqrt{2}} {\rm Tr} \left[ {\bar{B}}   (\xi M \xi + \xi^{\dagger}M^{\dagger} \xi^{\dagger})B \right] - \gamma \frac{F_\pi }{2 \sqrt{2}} {\rm Tr} \left[ {\bar{B}} \gamma_5  (\xi M \xi - \xi^{\dagger}M^{\dagger} \xi^{\dagger}) B \right] 
\end{eqnarray}
As done earlier we make explicit the relevant $U(1)$ axial phase via
\begin{equation}
\frac{F_\pi}{\sqrt{2}}\left( \xi^2 \right) _{ij} = U_{ij}= e^{- i \frac{\phi_i}{2}}  V_{ij}  e^{- i \frac{ \phi_i} {2}}   \ ,
\label{var67}
\end{equation}
implying
\begin{eqnarray}
 \xi_{ij} =  e^{- i \frac{\phi_i}{2}}  \nu_{i m} k_{mj}^{\dagger}   =k_{im}  \nu_{m j} e^{- i \frac{\phi_j}{2}} \ , \qquad {\rm with} \qquad  \nu = e^{ \frac{i\,\Phi}{\sqrt{2}F_\pi}} \ . 
\end{eqnarray}
Provided we transform the $B$ fields as in \eqref{Btransform} the previous Lagrangian becomes,
 \begin{eqnarray}
 L_{bar} & = & {\rm Tr} \left[{\bar{B}} i \gamma^{\mu} \partial_{\mu} B \right] -
\alpha \frac{F_\pi }{\sqrt{2}} {\rm Tr} \left[ {\bar{B}} B \right]  \nonumber \\
&&  + \frac{ F_{\pi} }{\sqrt{2}}  \delta \,\, {\rm Tr}  \left[ {\bar{B}}  B  M_p(\theta) +
  {\bar{B}}   \gamma_5 B  M_{m}(\theta) 
  \right] +  \frac{ F_{\pi} }{\sqrt{2}}  \gamma \,\,
  {\rm Tr}  \left[ {\bar{B}}    M_p(\theta) B -
  {\bar{B}}   \gamma_5   M_{m}(\theta) B   \right] 
\nonumber \\
&&+ a \left( \theta - \sum_{i} \phi_i \right)  \frac{ F_{\pi} }{\sqrt{2}}  {\rm Tr}   
\left[  \delta \left( \, {\bar{B}}  B \sin \left( \frac{\sqrt{2} }{F_{\pi}} \Phi \right) - i\,{\bar{B}}   \gamma_5 B \cos  \left( \frac{\sqrt{2} }{F_{\pi}} \Phi \right) \right) 
\right. \nonumber \\
&& \left. + \gamma \left( {\bar{B}}    \sin \left( \frac{\sqrt{2} }{F_{\pi}} \Phi \right) B  + i\,{\bar{B}}  \gamma_5  \cos  \left( \frac{\sqrt{2} }{F_{\pi}} \Phi \right) B   \right)  \right] \ ,
\label{efflag} 
\end{eqnarray}
with 
\begin{eqnarray}
M_{p/m} (\theta) \equiv \frac{\nu M(\theta) \nu \pm \nu^{\dagger} M(\theta) \nu^{\dagger}}{2}  \ . 
\end{eqnarray}
One can determine $\alpha, \gamma$ and $\delta$ in terms of the baryon
masses 
\begin{eqnarray}
\alpha = \frac{\sqrt{2}}{F_{\pi}} \left[ m_{\Sigma} + \frac{
    \mu^{2}}{( \mu_{3}^{2} - \mu^2)} ( 2 m_{\Sigma} - m_{\Xi} - m_N ) \right] \ ,
\label{alpha}
\end{eqnarray}
\begin{eqnarray}
\gamma = \frac{\sqrt{2}}{ F_{\pi}( \mu_{3}^{2} - \mu^2) } ( m_{\Sigma}- m_{\Xi} )  \ ,
\label{gamma}
\end{eqnarray}
\begin{eqnarray}
\delta = \frac{\sqrt{2}}{ F_{\pi}( \mu_{3}^{2} - \mu^2) } (m_{\Sigma} - m_{N})  \ .
\label{delta}
\end{eqnarray}
The baryon masses satisfy the Gell-Mann-Okubo
mass formula:
\begin{eqnarray}
3 m_{\Lambda} + m_{\Sigma} = 2 ( m_{\Xi} + m_{N} ) \ .
\label{gmo}
\end{eqnarray}
{From} the previous Lagrangian one can extract the $\pi N$ coupling
constants
\begin{eqnarray}
 {\bar{N}} \left[i \gamma_5 g_{\pi NN} + {\bar{g}}_{\pi NN} \,
\right] \pi^{i} \tau^{i} N \ .
\label{pinc}
\end{eqnarray}
The $CP$ violating one reads:
\begin{equation}
 {\bar{g}}_{\pi NN} = -\frac{a  ( \theta - \sum_{i} \phi_i )}{\mu^2_3 - \mu^2} \frac{m_\Xi - m_{\Sigma}}{F_\pi}  =-  \frac{\mu_{1}^{2}\mu_{2}^{2} \sin \theta}{ ({\mu^2_3 - \mu^2}) \sqrt{\mu_{1}^{4} +  \mu_{2}^{4} + 2\mu_{1}^{2} \mu_{2}^{2} \cos \theta }} \frac{m_\Xi - m_{\Sigma}}{F_\pi}\ . 
\end{equation}
In deriving the last identity we used Eq.~(\ref{cpvio}). We can also rewrite the previous expression in the chiral limit, directly in terms of the quark masses as:  
\begin{equation}
 {\bar{g}}_{\pi NN} = -  2\,\theta  \frac{\mu_{1}^{2}\mu_{2}^{2}  }{ ({2\mu^2_3 - \mu_1^2 - \mu_2^2})  (\mu_1^2 + \mu_2^2)} \frac{m_\Xi - m_{\Sigma}}{F_\pi} = - {2}\,\theta  \frac{m_{1}m_{2}  }{ ({2m_3 - m_1 - m_2})  (m_1 + m_2)} \frac{m_\Xi - m_{\Sigma}}{F_\pi}\ ,
\end{equation}
where we also assumed the small $\theta$ limit. 
For the $CP$ preserving coupling one must add new operators dictated by current algebra involving derivative couplings with the mesons. This leads to 
 \begin{eqnarray}
 F_{\pi} g_{\pi NN} \simeq m_N  \ .
\label{gt}
\end{eqnarray}
This is the Goldberger-Treiman relation (with $g_A =1$) apart from terms that vanish in the chiral limit.  Having computed ${\bar{g}}_{\pi  NN}$ we can use it to estimate the
electric dipole moment of the neutron that, if different from zero,
implies a violation of $CP$. The dominant contribution comes from the
two diagrams discussed and computed in Ref.~\cite{Crewther:1979pi} and one gets:
\begin{eqnarray}
D_n = \frac{1}{4 \pi^2 m_N} \cdot g_{\pi NN} {\bar{g}}_{\pi NN} \log
\frac{m_N}{m_{\pi}} = -  1.4 \cdot 10^{-15} \theta \, {\rm cm}  \ ,
\label{dn}
\end{eqnarray}
in units where the electric charge $e=1$. The experimental limit is:
\begin{eqnarray}
|D_n| < 6 \cdot 10^{-26} \ ,\qquad \Longrightarrow \qquad \theta < 10^{-10} \ .
\label{dnb}
\end{eqnarray}

\bibliographystyle{apsrev4-1}
\bibliography{biblio}

\begin{thebibliography}{90}%
\makeatletter
\providecommand \@ifxundefined [1]{%
 \@ifx{#1\undefined}
}%
\providecommand \@ifnum [1]{%
 \ifnum #1\expandafter \@firstoftwo
 \else \expandafter \@secondoftwo
 \fi
}%
\providecommand \@ifx [1]{%
 \ifx #1\expandafter \@firstoftwo
 \else \expandafter \@secondoftwo
 \fi
}%
\providecommand \natexlab [1]{#1}%
\providecommand \enquote  [1]{``#1''}%
\providecommand \bibnamefont  [1]{#1}%
\providecommand \bibfnamefont [1]{#1}%
\providecommand \citenamefont [1]{#1}%
\providecommand \href@noop [0]{\@secondoftwo}%
\providecommand \href [0]{\begingroup \@sanitize@url \@href}%
\providecommand \@href[1]{\@@startlink{#1}\@@href}%
\providecommand \@@href[1]{\endgroup#1\@@endlink}%
\providecommand \@sanitize@url [0]{\catcode `\\12\catcode `\$12\catcode
  `\&12\catcode `\#12\catcode `\^12\catcode `\_12\catcode `\%12\relax}%
\providecommand \@@startlink[1]{}%
\providecommand \@@endlink[0]{}%
\providecommand \url  [0]{\begingroup\@sanitize@url \@url }%
\providecommand \@url [1]{\endgroup\@href {#1}{\urlprefix }}%
\providecommand \urlprefix  [0]{URL }%
\providecommand \Eprint [0]{\href }%
\providecommand \doibase [0]{http://dx.doi.org/}%
\providecommand \selectlanguage [0]{\@gobble}%
\providecommand \bibinfo  [0]{\@secondoftwo}%
\providecommand \bibfield  [0]{\@secondoftwo}%
\providecommand \translation [1]{[#1]}%
\providecommand \BibitemOpen [0]{}%
\providecommand \bibitemStop [0]{}%
\providecommand \bibitemNoStop [0]{.\EOS\space}%
\providecommand \EOS [0]{\spacefactor3000\relax}%
\providecommand \BibitemShut  [1]{\csname bibitem#1\endcsname}%
\let\auto@bib@innerbib\@empty
\bibitem [{\citenamefont {Ade}\ \emph {et~al.}(2013)\citenamefont {Ade} \emph
  {et~al.}}]{Ade:2013zuv}%
  \BibitemOpen
  \bibfield  {author} {\bibinfo {author} {\bibfnamefont {P.}~\bibnamefont
  {Ade}} \emph {et~al.} (\bibinfo {collaboration} {Planck Collaboration}),\
  }\href@noop {} {\  (\bibinfo {year} {2013})},\ \Eprint
  {http://arxiv.org/abs/1303.5076} {arXiv:1303.5076 [astro-ph.CO]} \BibitemShut
  {NoStop}%
\bibitem [{\citenamefont {Crewther}\ \emph {et~al.}(1979)\citenamefont
  {Crewther}, \citenamefont {Di~Vecchia}, \citenamefont {Veneziano},\ and\
  \citenamefont {Witten}}]{Crewther:1979pi}%
  \BibitemOpen
  \bibfield  {author} {\bibinfo {author} {\bibfnamefont {R.}~\bibnamefont
  {Crewther}}, \bibinfo {author} {\bibfnamefont {P.}~\bibnamefont
  {Di~Vecchia}}, \bibinfo {author} {\bibfnamefont {G.}~\bibnamefont
  {Veneziano}}, \ and\ \bibinfo {author} {\bibfnamefont {E.}~\bibnamefont
  {Witten}},\ }\href {\doibase 10.1016/0370-2693(79)90128-X} {\bibfield
  {journal} {\bibinfo  {journal} {Phys.Lett.}\ }\textbf {\bibinfo {volume}
  {B88}},\ \bibinfo {pages} {123} (\bibinfo {year} {1979})}\BibitemShut
  {NoStop}%
\bibitem [{\citenamefont {Witten}(1979)}]{Witten:1979vv}%
  \BibitemOpen
  \bibfield  {author} {\bibinfo {author} {\bibfnamefont {E.}~\bibnamefont
  {Witten}},\ }\href {\doibase 10.1016/0550-3213(79)90031-2} {\bibfield
  {journal} {\bibinfo  {journal} {Nucl.Phys.}\ }\textbf {\bibinfo {volume}
  {B156}},\ \bibinfo {pages} {269} (\bibinfo {year} {1979})}\BibitemShut
  {NoStop}%
\bibitem [{\citenamefont {Veneziano}(1979)}]{Veneziano:1979ec}%
  \BibitemOpen
  \bibfield  {author} {\bibinfo {author} {\bibfnamefont {G.}~\bibnamefont
  {Veneziano}},\ }\href {\doibase 10.1016/0550-3213(79)90332-8} {\bibfield
  {journal} {\bibinfo  {journal} {Nucl.Phys.}\ }\textbf {\bibinfo {volume}
  {B159}},\ \bibinfo {pages} {213} (\bibinfo {year} {1979})}\BibitemShut
  {NoStop}%
\bibitem [{\citenamefont {Di~Vecchia}(1979)}]{DiVecchia:1979bf}%
  \BibitemOpen
  \bibfield  {author} {\bibinfo {author} {\bibfnamefont {P.}~\bibnamefont
  {Di~Vecchia}},\ }\href {\doibase 10.1016/0370-2693(79)91271-1} {\bibfield
  {journal} {\bibinfo  {journal} {Phys.Lett.}\ }\textbf {\bibinfo {volume}
  {B85}},\ \bibinfo {pages} {357} (\bibinfo {year} {1979})}\BibitemShut
  {NoStop}%
\bibitem [{\citenamefont {Rosenzweig}\ \emph {et~al.}(1980)\citenamefont
  {Rosenzweig}, \citenamefont {Schechter},\ and\ \citenamefont
  {Trahern}}]{Rosenzweig:1979ay}%
  \BibitemOpen
  \bibfield  {author} {\bibinfo {author} {\bibfnamefont {C.}~\bibnamefont
  {Rosenzweig}}, \bibinfo {author} {\bibfnamefont {J.}~\bibnamefont
  {Schechter}}, \ and\ \bibinfo {author} {\bibfnamefont {C.}~\bibnamefont
  {Trahern}},\ }\href {\doibase 10.1103/PhysRevD.21.3388} {\bibfield  {journal}
  {\bibinfo  {journal} {Phys.Rev.}\ }\textbf {\bibinfo {volume} {D21}},\
  \bibinfo {pages} {3388} (\bibinfo {year} {1980})}\BibitemShut {NoStop}%
\bibitem [{\citenamefont {Di~Vecchia}\ and\ \citenamefont
  {Veneziano}(1980{\natexlab{a}})}]{DiVecchia:1980ve}%
  \BibitemOpen
  \bibfield  {author} {\bibinfo {author} {\bibfnamefont {P.}~\bibnamefont
  {Di~Vecchia}}\ and\ \bibinfo {author} {\bibfnamefont {G.}~\bibnamefont
  {Veneziano}},\ }\href {\doibase 10.1016/0550-3213(80)90370-3} {\bibfield
  {journal} {\bibinfo  {journal} {Nucl.Phys.}\ }\textbf {\bibinfo {volume}
  {B171}},\ \bibinfo {pages} {253} (\bibinfo {year}
  {1980}{\natexlab{a}})}\BibitemShut {NoStop}%
\bibitem [{\citenamefont {Witten}(1980)}]{Witten:1980sp}%
  \BibitemOpen
  \bibfield  {author} {\bibinfo {author} {\bibfnamefont {E.}~\bibnamefont
  {Witten}},\ }\href {\doibase 10.1016/0003-4916(80)90325-5} {\bibfield
  {journal} {\bibinfo  {journal} {Annals Phys.}\ }\textbf {\bibinfo {volume}
  {128}},\ \bibinfo {pages} {363} (\bibinfo {year} {1980})}\BibitemShut
  {NoStop}%
\bibitem [{\citenamefont {Di~Vecchia}(1980)}]{DiVecchia:1980gi}%
  \BibitemOpen
  \bibfield  {author} {\bibinfo {author} {\bibfnamefont {P.}~\bibnamefont
  {Di~Vecchia}},\ }\href@noop {} {\bibfield  {journal} {\bibinfo  {journal}
  {Acta Physica Austriaca, Suppl. XXII, 341-381 (1980) by Springer-Verlag 1980.
  Lecture given at the XIX. Internationale Universitaetswochen fuer Kernphysik,
  Schladming, Austria, February 20-29, 1980.}\ } (\bibinfo {year}
  {1980})}\BibitemShut {NoStop}%
\bibitem [{\citenamefont {Narison}(2008)}]{Narison:2008jp}%
  \BibitemOpen
  \bibfield  {author} {\bibinfo {author} {\bibfnamefont {S.}~\bibnamefont
  {Narison}},\ }\href {\doibase 10.1016/j.physletb.2008.07.083} {\bibfield
  {journal} {\bibinfo  {journal} {Phys.Lett.}\ }\textbf {\bibinfo {volume}
  {B666}},\ \bibinfo {pages} {455} (\bibinfo {year} {2008})},\ \Eprint
  {http://arxiv.org/abs/0806.2618} {arXiv:0806.2618 [hep-ph]} \BibitemShut
  {NoStop}%
\bibitem [{\citenamefont {Beringer}\ \emph {et~al.}(2012)\citenamefont
  {Beringer} \emph {et~al.}}]{Beringer:1900zz}%
  \BibitemOpen
  \bibfield  {author} {\bibinfo {author} {\bibfnamefont {J.}~\bibnamefont
  {Beringer}} \emph {et~al.} (\bibinfo {collaboration} {Particle Data Group}),\
  }\href {\doibase 10.1103/PhysRevD.86.010001} {\bibfield  {journal} {\bibinfo
  {journal} {Phys.Rev.}\ }\textbf {\bibinfo {volume} {D86}},\ \bibinfo {pages}
  {010001} (\bibinfo {year} {2012})}\BibitemShut {NoStop}%
\bibitem [{\citenamefont {Aad}\ \emph {et~al.}(2012)\citenamefont {Aad} \emph
  {et~al.}}]{Aad:2012tfa}%
  \BibitemOpen
  \bibfield  {author} {\bibinfo {author} {\bibfnamefont {G.}~\bibnamefont
  {Aad}} \emph {et~al.} (\bibinfo {collaboration} {ATLAS Collaboration}),\
  }\href {\doibase 10.1016/j.physletb.2012.08.020} {\bibfield  {journal}
  {\bibinfo  {journal} {Phys.Lett.}\ }\textbf {\bibinfo {volume} {B716}},\
  \bibinfo {pages} {1} (\bibinfo {year} {2012})},\ \Eprint
  {http://arxiv.org/abs/1207.7214} {arXiv:1207.7214 [hep-ex]} \BibitemShut
  {NoStop}%
\bibitem [{\citenamefont {Chatrchyan}\ \emph {et~al.}(2012)\citenamefont
  {Chatrchyan} \emph {et~al.}}]{Chatrchyan:2012ufa}%
  \BibitemOpen
  \bibfield  {author} {\bibinfo {author} {\bibfnamefont {S.}~\bibnamefont
  {Chatrchyan}} \emph {et~al.} (\bibinfo {collaboration} {CMS Collaboration}),\
  }\href {\doibase 10.1016/j.physletb.2012.08.021} {\bibfield  {journal}
  {\bibinfo  {journal} {Phys.Lett.}\ }\textbf {\bibinfo {volume} {B716}},\
  \bibinfo {pages} {30} (\bibinfo {year} {2012})},\ \Eprint
  {http://arxiv.org/abs/1207.7235} {arXiv:1207.7235 [hep-ex]} \BibitemShut
  {NoStop}%
\bibitem [{\citenamefont {Guth}(1981)}]{Guth:1980zm}%
  \BibitemOpen
  \bibfield  {author} {\bibinfo {author} {\bibfnamefont {A.~H.}\ \bibnamefont
  {Guth}},\ }\href {\doibase 10.1103/PhysRevD.23.347} {\bibfield  {journal}
  {\bibinfo  {journal} {Phys.Rev.}\ }\textbf {\bibinfo {volume} {D23}},\
  \bibinfo {pages} {347} (\bibinfo {year} {1981})}\BibitemShut {NoStop}%
\bibitem [{\citenamefont {Linde}(1982)}]{Linde:1981mu}%
  \BibitemOpen
  \bibfield  {author} {\bibinfo {author} {\bibfnamefont {A.~D.}\ \bibnamefont
  {Linde}},\ }\href {\doibase 10.1016/0370-2693(82)91219-9} {\bibfield
  {journal} {\bibinfo  {journal} {Phys.Lett.}\ }\textbf {\bibinfo {volume}
  {B108}},\ \bibinfo {pages} {389} (\bibinfo {year} {1982})}\BibitemShut
  {NoStop}%
\bibitem [{\citenamefont {Sannino}\ and\ \citenamefont
  {Tuominen}(2005)}]{Sannino:2004qp}%
  \BibitemOpen
  \bibfield  {author} {\bibinfo {author} {\bibfnamefont {F.}~\bibnamefont
  {Sannino}}\ and\ \bibinfo {author} {\bibfnamefont {K.}~\bibnamefont
  {Tuominen}},\ }\href {\doibase 10.1103/PhysRevD.71.051901} {\bibfield
  {journal} {\bibinfo  {journal} {Phys.Rev.}\ }\textbf {\bibinfo {volume}
  {D71}},\ \bibinfo {pages} {051901} (\bibinfo {year} {2005})},\ \Eprint
  {http://arxiv.org/abs/hep-ph/0405209} {arXiv:hep-ph/0405209 [hep-ph]}
  \BibitemShut {NoStop}%
\bibitem [{\citenamefont {Dietrich}\ \emph {et~al.}(2005)\citenamefont
  {Dietrich}, \citenamefont {Sannino},\ and\ \citenamefont
  {Tuominen}}]{Dietrich:2005jn}%
  \BibitemOpen
  \bibfield  {author} {\bibinfo {author} {\bibfnamefont {D.~D.}\ \bibnamefont
  {Dietrich}}, \bibinfo {author} {\bibfnamefont {F.}~\bibnamefont {Sannino}}, \
  and\ \bibinfo {author} {\bibfnamefont {K.}~\bibnamefont {Tuominen}},\ }\href
  {\doibase 10.1103/PhysRevD.72.055001} {\bibfield  {journal} {\bibinfo
  {journal} {Phys.Rev.}\ }\textbf {\bibinfo {volume} {D72}},\ \bibinfo {pages}
  {055001} (\bibinfo {year} {2005})},\ \Eprint
  {http://arxiv.org/abs/hep-ph/0505059} {arXiv:hep-ph/0505059 [hep-ph]}
  \BibitemShut {NoStop}%
\bibitem [{\citenamefont {Dietrich}\ \emph {et~al.}(2006)\citenamefont
  {Dietrich}, \citenamefont {Sannino},\ and\ \citenamefont
  {Tuominen}}]{Dietrich:2005wk}%
  \BibitemOpen
  \bibfield  {author} {\bibinfo {author} {\bibfnamefont {D.~D.}\ \bibnamefont
  {Dietrich}}, \bibinfo {author} {\bibfnamefont {F.}~\bibnamefont {Sannino}}, \
  and\ \bibinfo {author} {\bibfnamefont {K.}~\bibnamefont {Tuominen}},\ }\href
  {\doibase 10.1103/PhysRevD.73.037701} {\bibfield  {journal} {\bibinfo
  {journal} {Phys.Rev.}\ }\textbf {\bibinfo {volume} {D73}},\ \bibinfo {pages}
  {037701} (\bibinfo {year} {2006})},\ \Eprint
  {http://arxiv.org/abs/hep-ph/0510217} {arXiv:hep-ph/0510217 [hep-ph]}
  \BibitemShut {NoStop}%
\bibitem [{\citenamefont {Dietrich}\ and\ \citenamefont
  {Sannino}(2007)}]{Dietrich:2006cm}%
  \BibitemOpen
  \bibfield  {author} {\bibinfo {author} {\bibfnamefont {D.~D.}\ \bibnamefont
  {Dietrich}}\ and\ \bibinfo {author} {\bibfnamefont {F.}~\bibnamefont
  {Sannino}},\ }\href {\doibase 10.1103/PhysRevD.75.085018} {\bibfield
  {journal} {\bibinfo  {journal} {Phys.Rev.}\ }\textbf {\bibinfo {volume}
  {D75}},\ \bibinfo {pages} {085018} (\bibinfo {year} {2007})},\ \Eprint
  {http://arxiv.org/abs/hep-ph/0611341} {arXiv:hep-ph/0611341 [hep-ph]}
  \BibitemShut {NoStop}%
\bibitem [{\citenamefont {Gudnason}\ \emph {et~al.}(2007)\citenamefont
  {Gudnason}, \citenamefont {Ryttov},\ and\ \citenamefont
  {Sannino}}]{Gudnason:2006mk}%
  \BibitemOpen
  \bibfield  {author} {\bibinfo {author} {\bibfnamefont {S.~B.}\ \bibnamefont
  {Gudnason}}, \bibinfo {author} {\bibfnamefont {T.~A.}\ \bibnamefont
  {Ryttov}}, \ and\ \bibinfo {author} {\bibfnamefont {F.}~\bibnamefont
  {Sannino}},\ }\href {\doibase 10.1103/PhysRevD.76.015005} {\bibfield
  {journal} {\bibinfo  {journal} {Phys.Rev.}\ }\textbf {\bibinfo {volume}
  {D76}},\ \bibinfo {pages} {015005} (\bibinfo {year} {2007})},\ \Eprint
  {http://arxiv.org/abs/hep-ph/0612230} {arXiv:hep-ph/0612230 [hep-ph]}
  \BibitemShut {NoStop}%
\bibitem [{\citenamefont {Ryttov}\ and\ \citenamefont
  {Sannino}(2008{\natexlab{a}})}]{Ryttov:2008xe}%
  \BibitemOpen
  \bibfield  {author} {\bibinfo {author} {\bibfnamefont {T.~A.}\ \bibnamefont
  {Ryttov}}\ and\ \bibinfo {author} {\bibfnamefont {F.}~\bibnamefont
  {Sannino}},\ }\href {\doibase 10.1103/PhysRevD.78.115010} {\bibfield
  {journal} {\bibinfo  {journal} {Phys.Rev.}\ }\textbf {\bibinfo {volume}
  {D78}},\ \bibinfo {pages} {115010} (\bibinfo {year} {2008}{\natexlab{a}})},\
  \Eprint {http://arxiv.org/abs/0809.0713} {arXiv:0809.0713 [hep-ph]}
  \BibitemShut {NoStop}%
\bibitem [{\citenamefont {Sannino}(2009{\natexlab{a}})}]{Sannino:2009aw}%
  \BibitemOpen
  \bibfield  {author} {\bibinfo {author} {\bibfnamefont {F.}~\bibnamefont
  {Sannino}},\ }\href {\doibase 10.1103/PhysRevD.79.096007} {\bibfield
  {journal} {\bibinfo  {journal} {Phys.Rev.}\ }\textbf {\bibinfo {volume}
  {D79}},\ \bibinfo {pages} {096007} (\bibinfo {year} {2009}{\natexlab{a}})},\
  \Eprint {http://arxiv.org/abs/0902.3494} {arXiv:0902.3494 [hep-ph]}
  \BibitemShut {NoStop}%
\bibitem [{\citenamefont {Frandsen}\ and\ \citenamefont
  {Sannino}(2010)}]{Frandsen:2009mi}%
  \BibitemOpen
  \bibfield  {author} {\bibinfo {author} {\bibfnamefont {M.~T.}\ \bibnamefont
  {Frandsen}}\ and\ \bibinfo {author} {\bibfnamefont {F.}~\bibnamefont
  {Sannino}},\ }\href {\doibase 10.1103/PhysRevD.81.097704} {\bibfield
  {journal} {\bibinfo  {journal} {Phys.Rev.}\ }\textbf {\bibinfo {volume}
  {D81}},\ \bibinfo {pages} {097704} (\bibinfo {year} {2010})},\ \Eprint
  {http://arxiv.org/abs/0911.1570} {arXiv:0911.1570 [hep-ph]} \BibitemShut
  {NoStop}%
\bibitem [{\citenamefont {Gudnason}\ \emph
  {et~al.}(2006{\natexlab{a}})\citenamefont {Gudnason}, \citenamefont
  {Kouvaris},\ and\ \citenamefont {Sannino}}]{Gudnason:2006ug}%
  \BibitemOpen
  \bibfield  {author} {\bibinfo {author} {\bibfnamefont {S.~B.}\ \bibnamefont
  {Gudnason}}, \bibinfo {author} {\bibfnamefont {C.}~\bibnamefont {Kouvaris}},
  \ and\ \bibinfo {author} {\bibfnamefont {F.}~\bibnamefont {Sannino}},\ }\href
  {\doibase 10.1103/PhysRevD.73.115003} {\bibfield  {journal} {\bibinfo
  {journal} {Phys.Rev.}\ }\textbf {\bibinfo {volume} {D73}},\ \bibinfo {pages}
  {115003} (\bibinfo {year} {2006}{\natexlab{a}})},\ \Eprint
  {http://arxiv.org/abs/hep-ph/0603014} {arXiv:hep-ph/0603014 [hep-ph]}
  \BibitemShut {NoStop}%
\bibitem [{\citenamefont {Gudnason}\ \emph
  {et~al.}(2006{\natexlab{b}})\citenamefont {Gudnason}, \citenamefont
  {Kouvaris},\ and\ \citenamefont {Sannino}}]{Gudnason:2006yj}%
  \BibitemOpen
  \bibfield  {author} {\bibinfo {author} {\bibfnamefont {S.~B.}\ \bibnamefont
  {Gudnason}}, \bibinfo {author} {\bibfnamefont {C.}~\bibnamefont {Kouvaris}},
  \ and\ \bibinfo {author} {\bibfnamefont {F.}~\bibnamefont {Sannino}},\ }\href
  {\doibase 10.1103/PhysRevD.74.095008} {\bibfield  {journal} {\bibinfo
  {journal} {Phys.Rev.}\ }\textbf {\bibinfo {volume} {D74}},\ \bibinfo {pages}
  {095008} (\bibinfo {year} {2006}{\natexlab{b}})},\ \Eprint
  {http://arxiv.org/abs/hep-ph/0608055} {arXiv:hep-ph/0608055 [hep-ph]}
  \BibitemShut {NoStop}%
\bibitem [{\citenamefont {Del~Nobile}\ \emph {et~al.}(2011)\citenamefont
  {Del~Nobile}, \citenamefont {Kouvaris},\ and\ \citenamefont
  {Sannino}}]{DelNobile:2011je}%
  \BibitemOpen
  \bibfield  {author} {\bibinfo {author} {\bibfnamefont {E.}~\bibnamefont
  {Del~Nobile}}, \bibinfo {author} {\bibfnamefont {C.}~\bibnamefont
  {Kouvaris}}, \ and\ \bibinfo {author} {\bibfnamefont {F.}~\bibnamefont
  {Sannino}},\ }\href {\doibase 10.1103/PhysRevD.84.027301} {\bibfield
  {journal} {\bibinfo  {journal} {Phys.Rev.}\ }\textbf {\bibinfo {volume}
  {D84}},\ \bibinfo {pages} {027301} (\bibinfo {year} {2011})},\ \Eprint
  {http://arxiv.org/abs/1105.5431} {arXiv:1105.5431 [hep-ph]} \BibitemShut
  {NoStop}%
\bibitem [{\citenamefont {Channuie}\ \emph {et~al.}(2011)\citenamefont
  {Channuie}, \citenamefont {Joergensen},\ and\ \citenamefont
  {Sannino}}]{Channuie:2011rq}%
  \BibitemOpen
  \bibfield  {author} {\bibinfo {author} {\bibfnamefont {P.}~\bibnamefont
  {Channuie}}, \bibinfo {author} {\bibfnamefont {J.~J.}\ \bibnamefont
  {Joergensen}}, \ and\ \bibinfo {author} {\bibfnamefont {F.}~\bibnamefont
  {Sannino}},\ }\href {\doibase 10.1088/1475-7516/2011/05/007} {\bibfield
  {journal} {\bibinfo  {journal} {JCAP}\ }\textbf {\bibinfo {volume} {1105}},\
  \bibinfo {pages} {007} (\bibinfo {year} {2011})},\ \Eprint
  {http://arxiv.org/abs/1102.2898} {arXiv:1102.2898 [hep-ph]} \BibitemShut
  {NoStop}%
\bibitem [{\citenamefont {Bezrukov}\ \emph {et~al.}(2012)\citenamefont
  {Bezrukov}, \citenamefont {Channuie}, \citenamefont {Joergensen},\ and\
  \citenamefont {Sannino}}]{Bezrukov:2011mv}%
  \BibitemOpen
  \bibfield  {author} {\bibinfo {author} {\bibfnamefont {F.}~\bibnamefont
  {Bezrukov}}, \bibinfo {author} {\bibfnamefont {P.}~\bibnamefont {Channuie}},
  \bibinfo {author} {\bibfnamefont {J.~J.}\ \bibnamefont {Joergensen}}, \ and\
  \bibinfo {author} {\bibfnamefont {F.}~\bibnamefont {Sannino}},\ }\href
  {\doibase 10.1103/PhysRevD.86.063513} {\bibfield  {journal} {\bibinfo
  {journal} {Phys.Rev.}\ }\textbf {\bibinfo {volume} {D86}},\ \bibinfo {pages}
  {063513} (\bibinfo {year} {2012})},\ \Eprint {http://arxiv.org/abs/1112.4054}
  {arXiv:1112.4054 [hep-ph]} \BibitemShut {NoStop}%
\bibitem [{\citenamefont {Channuie}\ \emph {et~al.}(2012)\citenamefont
  {Channuie}, \citenamefont {Jorgensen},\ and\ \citenamefont
  {Sannino}}]{Channuie:2012bv}%
  \BibitemOpen
  \bibfield  {author} {\bibinfo {author} {\bibfnamefont {P.}~\bibnamefont
  {Channuie}}, \bibinfo {author} {\bibfnamefont {J.~J.}\ \bibnamefont
  {Jorgensen}}, \ and\ \bibinfo {author} {\bibfnamefont {F.}~\bibnamefont
  {Sannino}},\ }\href {\doibase 10.1103/PhysRevD.86.125035} {\bibfield
  {journal} {\bibinfo  {journal} {Phys.Rev.}\ }\textbf {\bibinfo {volume}
  {D86}},\ \bibinfo {pages} {125035} (\bibinfo {year} {2012})},\ \Eprint
  {http://arxiv.org/abs/1209.6362} {arXiv:1209.6362 [hep-ph]} \BibitemShut
  {NoStop}%
\bibitem [{\citenamefont {Channuie}\ and\ \citenamefont
  {Karwan}(2013)}]{Channuie:2013lla}%
  \BibitemOpen
  \bibfield  {author} {\bibinfo {author} {\bibfnamefont {P.}~\bibnamefont
  {Channuie}}\ and\ \bibinfo {author} {\bibfnamefont {K.}~\bibnamefont
  {Karwan}},\ }\href@noop {} {\  (\bibinfo {year} {2013})},\ \Eprint
  {http://arxiv.org/abs/1307.2880} {arXiv:1307.2880 [hep-ph]} \BibitemShut
  {NoStop}%
\bibitem [{\citenamefont {Gell-Mann}\ \emph {et~al.}(1968)\citenamefont
  {Gell-Mann}, \citenamefont {Oakes},\ and\ \citenamefont
  {Renner}}]{GellMann:1968rz}%
  \BibitemOpen
  \bibfield  {author} {\bibinfo {author} {\bibfnamefont {M.}~\bibnamefont
  {Gell-Mann}}, \bibinfo {author} {\bibfnamefont {R.}~\bibnamefont {Oakes}}, \
  and\ \bibinfo {author} {\bibfnamefont {B.}~\bibnamefont {Renner}},\ }\href
  {\doibase 10.1103/PhysRev.175.2195} {\bibfield  {journal} {\bibinfo
  {journal} {Phys.Rev.}\ }\textbf {\bibinfo {volume} {175}},\ \bibinfo {pages}
  {2195} (\bibinfo {year} {1968})}\BibitemShut {NoStop}%
\bibitem [{\citenamefont {Peccei}\ and\ \citenamefont
  {Quinn}(1977{\natexlab{a}})}]{Peccei:1977hh}%
  \BibitemOpen
  \bibfield  {author} {\bibinfo {author} {\bibfnamefont {R.}~\bibnamefont
  {Peccei}}\ and\ \bibinfo {author} {\bibfnamefont {H.~R.}\ \bibnamefont
  {Quinn}},\ }\href {\doibase 10.1103/PhysRevLett.38.1440} {\bibfield
  {journal} {\bibinfo  {journal} {Phys.Rev.Lett.}\ }\textbf {\bibinfo {volume}
  {38}},\ \bibinfo {pages} {1440} (\bibinfo {year}
  {1977}{\natexlab{a}})}\BibitemShut {NoStop}%
\bibitem [{\citenamefont {Peccei}\ and\ \citenamefont
  {Quinn}(1977{\natexlab{b}})}]{Peccei:1977ur}%
  \BibitemOpen
  \bibfield  {author} {\bibinfo {author} {\bibfnamefont {R.}~\bibnamefont
  {Peccei}}\ and\ \bibinfo {author} {\bibfnamefont {H.~R.}\ \bibnamefont
  {Quinn}},\ }\href {\doibase 10.1103/PhysRevD.16.1791} {\bibfield  {journal}
  {\bibinfo  {journal} {Phys.Rev.}\ }\textbf {\bibinfo {volume} {D16}},\
  \bibinfo {pages} {1791} (\bibinfo {year} {1977}{\natexlab{b}})}\BibitemShut
  {NoStop}%
\bibitem [{\citenamefont {Sannino}(2009{\natexlab{b}})}]{Sannino:2009za}%
  \BibitemOpen
  \bibfield  {author} {\bibinfo {author} {\bibfnamefont {F.}~\bibnamefont
  {Sannino}},\ }\href@noop {} {\bibfield  {journal} {\bibinfo  {journal} {Acta
  Phys.Polon.}\ }\textbf {\bibinfo {volume} {B40}},\ \bibinfo {pages} {3533}
  (\bibinfo {year} {2009}{\natexlab{b}})},\ \Eprint
  {http://arxiv.org/abs/0911.0931} {arXiv:0911.0931 [hep-ph]} \BibitemShut
  {NoStop}%
\bibitem [{\citenamefont {Nussinov}(1985)}]{Nussinov:1985xr}%
  \BibitemOpen
  \bibfield  {author} {\bibinfo {author} {\bibfnamefont {S.}~\bibnamefont
  {Nussinov}},\ }\href {\doibase 10.1016/0370-2693(85)90689-6} {\bibfield
  {journal} {\bibinfo  {journal} {Phys.Lett.}\ }\textbf {\bibinfo {volume}
  {B165}},\ \bibinfo {pages} {55} (\bibinfo {year} {1985})}\BibitemShut
  {NoStop}%
\bibitem [{\citenamefont {Barr}\ \emph {et~al.}(1990)\citenamefont {Barr},
  \citenamefont {Chivukula},\ and\ \citenamefont {Farhi}}]{Barr:1990ca}%
  \BibitemOpen
  \bibfield  {author} {\bibinfo {author} {\bibfnamefont {S.~M.}\ \bibnamefont
  {Barr}}, \bibinfo {author} {\bibfnamefont {R.~S.}\ \bibnamefont {Chivukula}},
  \ and\ \bibinfo {author} {\bibfnamefont {E.}~\bibnamefont {Farhi}},\ }\href
  {\doibase 10.1016/0370-2693(90)91661-T} {\bibfield  {journal} {\bibinfo
  {journal} {Phys.Lett.}\ }\textbf {\bibinfo {volume} {B241}},\ \bibinfo
  {pages} {387} (\bibinfo {year} {1990})}\BibitemShut {NoStop}%
\bibitem [{\citenamefont {Foadi}\ \emph {et~al.}(2009)\citenamefont {Foadi},
  \citenamefont {Frandsen},\ and\ \citenamefont {Sannino}}]{Foadi:2008qv}%
  \BibitemOpen
  \bibfield  {author} {\bibinfo {author} {\bibfnamefont {R.}~\bibnamefont
  {Foadi}}, \bibinfo {author} {\bibfnamefont {M.~T.}\ \bibnamefont {Frandsen}},
  \ and\ \bibinfo {author} {\bibfnamefont {F.}~\bibnamefont {Sannino}},\ }\href
  {\doibase 10.1103/PhysRevD.80.037702} {\bibfield  {journal} {\bibinfo
  {journal} {Phys.Rev.}\ }\textbf {\bibinfo {volume} {D80}},\ \bibinfo {pages}
  {037702} (\bibinfo {year} {2009})},\ \Eprint {http://arxiv.org/abs/0812.3406}
  {arXiv:0812.3406 [hep-ph]} \BibitemShut {NoStop}%
\bibitem [{\citenamefont {Nardi}\ \emph {et~al.}(2009)\citenamefont {Nardi},
  \citenamefont {Sannino},\ and\ \citenamefont {Strumia}}]{Nardi:2008ix}%
  \BibitemOpen
  \bibfield  {author} {\bibinfo {author} {\bibfnamefont {E.}~\bibnamefont
  {Nardi}}, \bibinfo {author} {\bibfnamefont {F.}~\bibnamefont {Sannino}}, \
  and\ \bibinfo {author} {\bibfnamefont {A.}~\bibnamefont {Strumia}},\ }\href
  {\doibase 10.1088/1475-7516/2009/01/043} {\bibfield  {journal} {\bibinfo
  {journal} {JCAP}\ }\textbf {\bibinfo {volume} {0901}},\ \bibinfo {pages}
  {043} (\bibinfo {year} {2009})},\ \Eprint {http://arxiv.org/abs/0811.4153}
  {arXiv:0811.4153 [hep-ph]} \BibitemShut {NoStop}%
\bibitem [{\citenamefont {Sannino}\ and\ \citenamefont
  {Zwicky}(2009)}]{Sannino:2008nv}%
  \BibitemOpen
  \bibfield  {author} {\bibinfo {author} {\bibfnamefont {F.}~\bibnamefont
  {Sannino}}\ and\ \bibinfo {author} {\bibfnamefont {R.}~\bibnamefont
  {Zwicky}},\ }\href {\doibase 10.1103/PhysRevD.79.015016} {\bibfield
  {journal} {\bibinfo  {journal} {Phys.Rev.}\ }\textbf {\bibinfo {volume}
  {D79}},\ \bibinfo {pages} {015016} (\bibinfo {year} {2009})},\ \Eprint
  {http://arxiv.org/abs/0810.2686} {arXiv:0810.2686 [hep-ph]} \BibitemShut
  {NoStop}%
\bibitem [{\citenamefont {Buckley}\ and\ \citenamefont
  {Neil}(2013)}]{Buckley:2012ky}%
  \BibitemOpen
  \bibfield  {author} {\bibinfo {author} {\bibfnamefont {M.~R.}\ \bibnamefont
  {Buckley}}\ and\ \bibinfo {author} {\bibfnamefont {E.~T.}\ \bibnamefont
  {Neil}},\ }\href {\doibase 10.1103/PhysRevD.87.043510} {\bibfield  {journal}
  {\bibinfo  {journal} {Phys.Rev.}\ }\textbf {\bibinfo {volume} {D87}},\
  \bibinfo {pages} {043510} (\bibinfo {year} {2013})},\ \Eprint
  {http://arxiv.org/abs/1209.6054} {arXiv:1209.6054 [hep-ph]} \BibitemShut
  {NoStop}%
\bibitem [{\citenamefont {Lewis}\ \emph {et~al.}(2012)\citenamefont {Lewis},
  \citenamefont {Pica},\ and\ \citenamefont {Sannino}}]{Lewis:2011zb}%
  \BibitemOpen
  \bibfield  {author} {\bibinfo {author} {\bibfnamefont {R.}~\bibnamefont
  {Lewis}}, \bibinfo {author} {\bibfnamefont {C.}~\bibnamefont {Pica}}, \ and\
  \bibinfo {author} {\bibfnamefont {F.}~\bibnamefont {Sannino}},\ }\href
  {\doibase 10.1103/PhysRevD.85.014504} {\bibfield  {journal} {\bibinfo
  {journal} {Phys.Rev.}\ }\textbf {\bibinfo {volume} {D85}},\ \bibinfo {pages}
  {014504} (\bibinfo {year} {2012})},\ \Eprint {http://arxiv.org/abs/1109.3513}
  {arXiv:1109.3513 [hep-ph]} \BibitemShut {NoStop}%
\bibitem [{\citenamefont {Hietanen}\ \emph
  {et~al.}(2013{\natexlab{a}})\citenamefont {Hietanen}, \citenamefont {Pica},
  \citenamefont {Sannino},\ and\ \citenamefont
  {Sondergaard}}]{Hietanen:2012sz}%
  \BibitemOpen
  \bibfield  {author} {\bibinfo {author} {\bibfnamefont {A.}~\bibnamefont
  {Hietanen}}, \bibinfo {author} {\bibfnamefont {C.}~\bibnamefont {Pica}},
  \bibinfo {author} {\bibfnamefont {F.}~\bibnamefont {Sannino}}, \ and\
  \bibinfo {author} {\bibfnamefont {U.~I.}\ \bibnamefont {Sondergaard}},\
  }\href {\doibase 10.1103/PhysRevD.87.034508} {\bibfield  {journal} {\bibinfo
  {journal} {Phys.Rev.}\ }\textbf {\bibinfo {volume} {D87}},\ \bibinfo {pages}
  {034508} (\bibinfo {year} {2013}{\natexlab{a}})},\ \Eprint
  {http://arxiv.org/abs/1211.5021} {arXiv:1211.5021 [hep-lat]} \BibitemShut
  {NoStop}%
\bibitem [{\citenamefont {Appelquist}\ \emph {et~al.}(2013)\citenamefont
  {Appelquist}, \citenamefont {Brower}, \citenamefont {Buchoff}, \citenamefont
  {Cheng}, \citenamefont {Cohen} \emph {et~al.}}]{Appelquist:2013ms}%
  \BibitemOpen
  \bibfield  {author} {\bibinfo {author} {\bibfnamefont {T.}~\bibnamefont
  {Appelquist}}, \bibinfo {author} {\bibfnamefont {R.}~\bibnamefont {Brower}},
  \bibinfo {author} {\bibfnamefont {M.}~\bibnamefont {Buchoff}}, \bibinfo
  {author} {\bibfnamefont {M.}~\bibnamefont {Cheng}}, \bibinfo {author}
  {\bibfnamefont {S.}~\bibnamefont {Cohen}},  \emph {et~al.},\ }\href {\doibase
  10.1103/PhysRevD.88.014502} {\bibfield  {journal} {\bibinfo  {journal}
  {Phys.Rev.}\ }\textbf {\bibinfo {volume} {D88}},\ \bibinfo {pages} {014502}
  (\bibinfo {year} {2013})},\ \Eprint {http://arxiv.org/abs/1301.1693}
  {arXiv:1301.1693 [hep-ph]} \BibitemShut {NoStop}%
\bibitem [{\citenamefont {Hietanen}\ \emph
  {et~al.}(2013{\natexlab{b}})\citenamefont {Hietanen}, \citenamefont {Lewis},
  \citenamefont {Pica},\ and\ \citenamefont {Sannino}}]{Hietanen:2013fya}%
  \BibitemOpen
  \bibfield  {author} {\bibinfo {author} {\bibfnamefont {A.}~\bibnamefont
  {Hietanen}}, \bibinfo {author} {\bibfnamefont {R.}~\bibnamefont {Lewis}},
  \bibinfo {author} {\bibfnamefont {C.}~\bibnamefont {Pica}}, \ and\ \bibinfo
  {author} {\bibfnamefont {F.}~\bibnamefont {Sannino}},\ }\href@noop {} {\
  (\bibinfo {year} {2013}{\natexlab{b}})},\ \Eprint
  {http://arxiv.org/abs/1308.4130} {arXiv:1308.4130 [hep-ph]} \BibitemShut
  {NoStop}%
\bibitem [{\citenamefont {Di~Vecchia}\ and\ \citenamefont
  {Veneziano}(1980{\natexlab{b}})}]{DiVecchia:1980xq}%
  \BibitemOpen
  \bibfield  {author} {\bibinfo {author} {\bibfnamefont {P.}~\bibnamefont
  {Di~Vecchia}}\ and\ \bibinfo {author} {\bibfnamefont {G.}~\bibnamefont
  {Veneziano}},\ }\href {\doibase 10.1016/0370-2693(80)90480-3} {\bibfield
  {journal} {\bibinfo  {journal} {Phys.Lett.}\ }\textbf {\bibinfo {volume}
  {B95}},\ \bibinfo {pages} {247} (\bibinfo {year}
  {1980}{\natexlab{b}})}\BibitemShut {NoStop}%
\bibitem [{\citenamefont {Corrigan}\ and\ \citenamefont
  {Ramond}(1979)}]{Corrigan:1979xf}%
  \BibitemOpen
  \bibfield  {author} {\bibinfo {author} {\bibfnamefont {E.}~\bibnamefont
  {Corrigan}}\ and\ \bibinfo {author} {\bibfnamefont {P.}~\bibnamefont
  {Ramond}},\ }\href {\doibase 10.1016/0370-2693(79)90022-4} {\bibfield
  {journal} {\bibinfo  {journal} {Phys.Lett.}\ }\textbf {\bibinfo {volume}
  {B87}},\ \bibinfo {pages} {73} (\bibinfo {year} {1979})}\BibitemShut
  {NoStop}%
\bibitem [{\citenamefont {Kiritsis}\ and\ \citenamefont
  {Papavassiliou}(1990)}]{Kiritsis:1989ge}%
  \BibitemOpen
  \bibfield  {author} {\bibinfo {author} {\bibfnamefont {E.~B.}\ \bibnamefont
  {Kiritsis}}\ and\ \bibinfo {author} {\bibfnamefont {J.}~\bibnamefont
  {Papavassiliou}},\ }\href {\doibase 10.1103/PhysRevD.42.4238} {\bibfield
  {journal} {\bibinfo  {journal} {Phys.Rev.}\ }\textbf {\bibinfo {volume}
  {D42}},\ \bibinfo {pages} {4238} (\bibinfo {year} {1990})}\BibitemShut
  {NoStop}%
\bibitem [{\citenamefont {Armoni}\ \emph {et~al.}(2003)\citenamefont {Armoni},
  \citenamefont {Shifman},\ and\ \citenamefont {Veneziano}}]{Armoni:2003gp}%
  \BibitemOpen
  \bibfield  {author} {\bibinfo {author} {\bibfnamefont {A.}~\bibnamefont
  {Armoni}}, \bibinfo {author} {\bibfnamefont {M.}~\bibnamefont {Shifman}}, \
  and\ \bibinfo {author} {\bibfnamefont {G.}~\bibnamefont {Veneziano}},\ }\href
  {\doibase 10.1016/S0550-3213(03)00538-8} {\bibfield  {journal} {\bibinfo
  {journal} {Nucl.Phys.}\ }\textbf {\bibinfo {volume} {B667}},\ \bibinfo
  {pages} {170} (\bibinfo {year} {2003})}\BibitemShut {NoStop}%
\bibitem [{\citenamefont {Sannino}\ and\ \citenamefont
  {Shifman}(2004)}]{Sannino:2003xe}%
  \BibitemOpen
  \bibfield  {author} {\bibinfo {author} {\bibfnamefont {F.}~\bibnamefont
  {Sannino}}\ and\ \bibinfo {author} {\bibfnamefont {M.}~\bibnamefont
  {Shifman}},\ }\href {\doibase 10.1103/PhysRevD.69.125004} {\bibfield
  {journal} {\bibinfo  {journal} {Phys.Rev.}\ }\textbf {\bibinfo {volume}
  {D69}},\ \bibinfo {pages} {125004} (\bibinfo {year} {2004})},\ \Eprint
  {http://arxiv.org/abs/hep-th/0309252} {arXiv:hep-th/0309252 [hep-th]}
  \BibitemShut {NoStop}%
\bibitem [{\citenamefont {Sannino}\ and\ \citenamefont
  {Schechter}(2007)}]{Sannino:2007yp}%
  \BibitemOpen
  \bibfield  {author} {\bibinfo {author} {\bibfnamefont {F.}~\bibnamefont
  {Sannino}}\ and\ \bibinfo {author} {\bibfnamefont {J.}~\bibnamefont
  {Schechter}},\ }\href {\doibase 10.1103/PhysRevD.76.014014} {\bibfield
  {journal} {\bibinfo  {journal} {Phys.Rev.}\ }\textbf {\bibinfo {volume}
  {D76}},\ \bibinfo {pages} {014014} (\bibinfo {year} {2007})},\ \Eprint
  {http://arxiv.org/abs/0704.0602} {arXiv:0704.0602 [hep-ph]} \BibitemShut
  {NoStop}%
\bibitem [{\citenamefont {Hong}\ \emph {et~al.}(2004)\citenamefont {Hong},
  \citenamefont {Hsu},\ and\ \citenamefont {Sannino}}]{Hong:2004td}%
  \BibitemOpen
  \bibfield  {author} {\bibinfo {author} {\bibfnamefont {D.~K.}\ \bibnamefont
  {Hong}}, \bibinfo {author} {\bibfnamefont {S.~D.}\ \bibnamefont {Hsu}}, \
  and\ \bibinfo {author} {\bibfnamefont {F.}~\bibnamefont {Sannino}},\ }\href
  {\doibase 10.1016/j.physletb.2004.07.007} {\bibfield  {journal} {\bibinfo
  {journal} {Phys.Lett.}\ }\textbf {\bibinfo {volume} {B597}},\ \bibinfo
  {pages} {89} (\bibinfo {year} {2004})},\ \Eprint
  {http://arxiv.org/abs/hep-ph/0406200} {arXiv:hep-ph/0406200 [hep-ph]}
  \BibitemShut {NoStop}%
\bibitem [{\citenamefont {Catterall}\ and\ \citenamefont
  {Sannino}(2007)}]{Catterall:2007yx}%
  \BibitemOpen
  \bibfield  {author} {\bibinfo {author} {\bibfnamefont {S.}~\bibnamefont
  {Catterall}}\ and\ \bibinfo {author} {\bibfnamefont {F.}~\bibnamefont
  {Sannino}},\ }\href {\doibase 10.1103/PhysRevD.76.034504} {\bibfield
  {journal} {\bibinfo  {journal} {Phys.Rev.}\ }\textbf {\bibinfo {volume}
  {D76}},\ \bibinfo {pages} {034504} (\bibinfo {year} {2007})},\ \Eprint
  {http://arxiv.org/abs/0705.1664} {arXiv:0705.1664 [hep-lat]} \BibitemShut
  {NoStop}%
\bibitem [{\citenamefont {Catterall}\ \emph {et~al.}(2008)\citenamefont
  {Catterall}, \citenamefont {Giedt}, \citenamefont {Sannino},\ and\
  \citenamefont {Schneible}}]{Catterall:2008qk}%
  \BibitemOpen
  \bibfield  {author} {\bibinfo {author} {\bibfnamefont {S.}~\bibnamefont
  {Catterall}}, \bibinfo {author} {\bibfnamefont {J.}~\bibnamefont {Giedt}},
  \bibinfo {author} {\bibfnamefont {F.}~\bibnamefont {Sannino}}, \ and\
  \bibinfo {author} {\bibfnamefont {J.}~\bibnamefont {Schneible}},\ }\href
  {\doibase 10.1088/1126-6708/2008/11/009} {\bibfield  {journal} {\bibinfo
  {journal} {JHEP}\ }\textbf {\bibinfo {volume} {0811}},\ \bibinfo {pages}
  {009} (\bibinfo {year} {2008})},\ \Eprint {http://arxiv.org/abs/0807.0792}
  {arXiv:0807.0792 [hep-lat]} \BibitemShut {NoStop}%
\bibitem [{\citenamefont {Hietanen}\ \emph {et~al.}(2009)\citenamefont
  {Hietanen}, \citenamefont {Rantaharju}, \citenamefont {Rummukainen},\ and\
  \citenamefont {Tuominen}}]{Hietanen:2008mr}%
  \BibitemOpen
  \bibfield  {author} {\bibinfo {author} {\bibfnamefont {A.~J.}\ \bibnamefont
  {Hietanen}}, \bibinfo {author} {\bibfnamefont {J.}~\bibnamefont
  {Rantaharju}}, \bibinfo {author} {\bibfnamefont {K.}~\bibnamefont
  {Rummukainen}}, \ and\ \bibinfo {author} {\bibfnamefont {K.}~\bibnamefont
  {Tuominen}},\ }\href {\doibase 10.1088/1126-6708/2009/05/025} {\bibfield
  {journal} {\bibinfo  {journal} {JHEP}\ }\textbf {\bibinfo {volume} {0905}},\
  \bibinfo {pages} {025} (\bibinfo {year} {2009})},\ \Eprint
  {http://arxiv.org/abs/0812.1467} {arXiv:0812.1467 [hep-lat]} \BibitemShut
  {NoStop}%
\bibitem [{\citenamefont {Catterall}\ \emph {et~al.}(2009)\citenamefont
  {Catterall}, \citenamefont {Giedt}, \citenamefont {Sannino},\ and\
  \citenamefont {Schneible}}]{Catterall:2009sb}%
  \BibitemOpen
  \bibfield  {author} {\bibinfo {author} {\bibfnamefont {S.}~\bibnamefont
  {Catterall}}, \bibinfo {author} {\bibfnamefont {J.}~\bibnamefont {Giedt}},
  \bibinfo {author} {\bibfnamefont {F.}~\bibnamefont {Sannino}}, \ and\
  \bibinfo {author} {\bibfnamefont {J.}~\bibnamefont {Schneible}},\ }\href@noop
  {} {\  (\bibinfo {year} {2009})},\ \Eprint {http://arxiv.org/abs/0910.4387}
  {arXiv:0910.4387 [hep-lat]} \BibitemShut {NoStop}%
\bibitem [{\citenamefont {Del~Debbio}\ \emph {et~al.}(2010)\citenamefont
  {Del~Debbio}, \citenamefont {Lucini}, \citenamefont {Patella}, \citenamefont
  {Pica},\ and\ \citenamefont {Rago}}]{DelDebbio:2010hx}%
  \BibitemOpen
  \bibfield  {author} {\bibinfo {author} {\bibfnamefont {L.}~\bibnamefont
  {Del~Debbio}}, \bibinfo {author} {\bibfnamefont {B.}~\bibnamefont {Lucini}},
  \bibinfo {author} {\bibfnamefont {A.}~\bibnamefont {Patella}}, \bibinfo
  {author} {\bibfnamefont {C.}~\bibnamefont {Pica}}, \ and\ \bibinfo {author}
  {\bibfnamefont {A.}~\bibnamefont {Rago}},\ }\href {\doibase
  10.1103/PhysRevD.82.014510} {\bibfield  {journal} {\bibinfo  {journal}
  {Phys.Rev.}\ }\textbf {\bibinfo {volume} {D82}},\ \bibinfo {pages} {014510}
  (\bibinfo {year} {2010})},\ \Eprint {http://arxiv.org/abs/1004.3206}
  {arXiv:1004.3206 [hep-lat]} \BibitemShut {NoStop}%
\bibitem [{\citenamefont {Kogut}\ and\ \citenamefont
  {Sinclair}(2010)}]{Kogut:2010cz}%
  \BibitemOpen
  \bibfield  {author} {\bibinfo {author} {\bibfnamefont {J.}~\bibnamefont
  {Kogut}}\ and\ \bibinfo {author} {\bibfnamefont {D.}~\bibnamefont
  {Sinclair}},\ }\href {\doibase 10.1103/PhysRevD.81.114507} {\bibfield
  {journal} {\bibinfo  {journal} {Phys.Rev.}\ }\textbf {\bibinfo {volume}
  {D81}},\ \bibinfo {pages} {114507} (\bibinfo {year} {2010})},\ \Eprint
  {http://arxiv.org/abs/1002.2988} {arXiv:1002.2988 [hep-lat]} \BibitemShut
  {NoStop}%
\bibitem [{\citenamefont {Kogut}\ and\ \citenamefont
  {Sinclair}(2012)}]{Kogut:2011bd}%
  \BibitemOpen
  \bibfield  {author} {\bibinfo {author} {\bibfnamefont {J.}~\bibnamefont
  {Kogut}}\ and\ \bibinfo {author} {\bibfnamefont {D.}~\bibnamefont
  {Sinclair}},\ }\href {\doibase 10.1103/PhysRevD.85.054505} {\bibfield
  {journal} {\bibinfo  {journal} {Phys.Rev.}\ }\textbf {\bibinfo {volume}
  {D85}},\ \bibinfo {pages} {054505} (\bibinfo {year} {2012})},\ \Eprint
  {http://arxiv.org/abs/1111.3353} {arXiv:1111.3353 [hep-lat]} \BibitemShut
  {NoStop}%
\bibitem [{\citenamefont {Bursa}\ \emph {et~al.}(2011)\citenamefont {Bursa},
  \citenamefont {Del~Debbio}, \citenamefont {Henty}, \citenamefont {Kerrane},
  \citenamefont {Lucini} \emph {et~al.}}]{Bursa:2011ru}%
  \BibitemOpen
  \bibfield  {author} {\bibinfo {author} {\bibfnamefont {F.}~\bibnamefont
  {Bursa}}, \bibinfo {author} {\bibfnamefont {L.}~\bibnamefont {Del~Debbio}},
  \bibinfo {author} {\bibfnamefont {D.}~\bibnamefont {Henty}}, \bibinfo
  {author} {\bibfnamefont {E.}~\bibnamefont {Kerrane}}, \bibinfo {author}
  {\bibfnamefont {B.}~\bibnamefont {Lucini}},  \emph {et~al.},\ }\href
  {\doibase 10.1103/PhysRevD.84.034506} {\bibfield  {journal} {\bibinfo
  {journal} {Phys.Rev.}\ }\textbf {\bibinfo {volume} {D84}},\ \bibinfo {pages}
  {034506} (\bibinfo {year} {2011})},\ \Eprint {http://arxiv.org/abs/1104.4301}
  {arXiv:1104.4301 [hep-lat]} \BibitemShut {NoStop}%
\bibitem [{\citenamefont {Giedt}\ and\ \citenamefont
  {Weinberg}(2012)}]{Giedt:2012rj}%
  \BibitemOpen
  \bibfield  {author} {\bibinfo {author} {\bibfnamefont {J.}~\bibnamefont
  {Giedt}}\ and\ \bibinfo {author} {\bibfnamefont {E.}~\bibnamefont
  {Weinberg}},\ }\href {\doibase 10.1103/PhysRevD.85.097503} {\bibfield
  {journal} {\bibinfo  {journal} {Phys.Rev.}\ }\textbf {\bibinfo {volume}
  {D85}},\ \bibinfo {pages} {097503} (\bibinfo {year} {2012})},\ \Eprint
  {http://arxiv.org/abs/1201.6262} {arXiv:1201.6262 [hep-lat]} \BibitemShut
  {NoStop}%
\bibitem [{\citenamefont {Fodor}\ \emph {et~al.}(2012)\citenamefont {Fodor},
  \citenamefont {Holland}, \citenamefont {Kuti}, \citenamefont {Nogradi},
  \citenamefont {Schroeder} \emph {et~al.}}]{Fodor:2012ty}%
  \BibitemOpen
  \bibfield  {author} {\bibinfo {author} {\bibfnamefont {Z.}~\bibnamefont
  {Fodor}}, \bibinfo {author} {\bibfnamefont {K.}~\bibnamefont {Holland}},
  \bibinfo {author} {\bibfnamefont {J.}~\bibnamefont {Kuti}}, \bibinfo {author}
  {\bibfnamefont {D.}~\bibnamefont {Nogradi}}, \bibinfo {author} {\bibfnamefont
  {C.}~\bibnamefont {Schroeder}},  \emph {et~al.},\ }\href {\doibase
  10.1016/j.physletb.2012.10.079} {\bibfield  {journal} {\bibinfo  {journal}
  {Phys.Lett.}\ }\textbf {\bibinfo {volume} {B718}},\ \bibinfo {pages} {657}
  (\bibinfo {year} {2012})},\ \Eprint {http://arxiv.org/abs/1209.0391}
  {arXiv:1209.0391 [hep-lat]} \BibitemShut {NoStop}%
\bibitem [{\citenamefont {Appelquist}\ and\ \citenamefont
  {Sannino}(1999)}]{Appelquist:1998xf}%
  \BibitemOpen
  \bibfield  {author} {\bibinfo {author} {\bibfnamefont {T.}~\bibnamefont
  {Appelquist}}\ and\ \bibinfo {author} {\bibfnamefont {F.}~\bibnamefont
  {Sannino}},\ }\href {\doibase 10.1103/PhysRevD.59.067702} {\bibfield
  {journal} {\bibinfo  {journal} {Phys.Rev.}\ }\textbf {\bibinfo {volume}
  {D59}},\ \bibinfo {pages} {067702} (\bibinfo {year} {1999})},\ \Eprint
  {http://arxiv.org/abs/hep-ph/9806409} {arXiv:hep-ph/9806409 [hep-ph]}
  \BibitemShut {NoStop}%
\bibitem [{\citenamefont {Foadi}\ \emph {et~al.}(2007)\citenamefont {Foadi},
  \citenamefont {Frandsen}, \citenamefont {Ryttov},\ and\ \citenamefont
  {Sannino}}]{Foadi:2007ue}%
  \BibitemOpen
  \bibfield  {author} {\bibinfo {author} {\bibfnamefont {R.}~\bibnamefont
  {Foadi}}, \bibinfo {author} {\bibfnamefont {M.~T.}\ \bibnamefont {Frandsen}},
  \bibinfo {author} {\bibfnamefont {T.~A.}\ \bibnamefont {Ryttov}}, \ and\
  \bibinfo {author} {\bibfnamefont {F.}~\bibnamefont {Sannino}},\ }\href
  {\doibase 10.1103/PhysRevD.76.055005} {\bibfield  {journal} {\bibinfo
  {journal} {Phys.Rev.}\ }\textbf {\bibinfo {volume} {D76}},\ \bibinfo {pages}
  {055005} (\bibinfo {year} {2007})},\ \Eprint {http://arxiv.org/abs/0706.1696}
  {arXiv:0706.1696 [hep-ph]} \BibitemShut {NoStop}%
\bibitem [{\citenamefont {Belyaev}\ \emph {et~al.}(2009)\citenamefont
  {Belyaev}, \citenamefont {Foadi}, \citenamefont {Frandsen}, \citenamefont
  {Jarvinen}, \citenamefont {Sannino} \emph {et~al.}}]{Belyaev:2008yj}%
  \BibitemOpen
  \bibfield  {author} {\bibinfo {author} {\bibfnamefont {A.}~\bibnamefont
  {Belyaev}}, \bibinfo {author} {\bibfnamefont {R.}~\bibnamefont {Foadi}},
  \bibinfo {author} {\bibfnamefont {M.~T.}\ \bibnamefont {Frandsen}}, \bibinfo
  {author} {\bibfnamefont {M.}~\bibnamefont {Jarvinen}}, \bibinfo {author}
  {\bibfnamefont {F.}~\bibnamefont {Sannino}},  \emph {et~al.},\ }\href
  {\doibase 10.1103/PhysRevD.79.035006} {\bibfield  {journal} {\bibinfo
  {journal} {Phys.Rev.}\ }\textbf {\bibinfo {volume} {D79}},\ \bibinfo {pages}
  {035006} (\bibinfo {year} {2009})},\ \Eprint {http://arxiv.org/abs/0809.0793}
  {arXiv:0809.0793 [hep-ph]} \BibitemShut {NoStop}%
\bibitem [{\citenamefont {Andersen}\ \emph {et~al.}(2011)\citenamefont
  {Andersen}, \citenamefont {Antipin}, \citenamefont {Azuelos}, \citenamefont
  {Del~Debbio}, \citenamefont {Del~Nobile} \emph {et~al.}}]{Andersen:2011yj}%
  \BibitemOpen
  \bibfield  {author} {\bibinfo {author} {\bibfnamefont {J.}~\bibnamefont
  {Andersen}}, \bibinfo {author} {\bibfnamefont {O.}~\bibnamefont {Antipin}},
  \bibinfo {author} {\bibfnamefont {G.}~\bibnamefont {Azuelos}}, \bibinfo
  {author} {\bibfnamefont {L.}~\bibnamefont {Del~Debbio}}, \bibinfo {author}
  {\bibfnamefont {E.}~\bibnamefont {Del~Nobile}},  \emph {et~al.},\ }\href
  {\doibase 10.1140/epjp/i2011-11081-1} {\bibfield  {journal} {\bibinfo
  {journal} {Eur.Phys.J.Plus}\ }\textbf {\bibinfo {volume} {126}},\ \bibinfo
  {pages} {81} (\bibinfo {year} {2011})},\ \Eprint
  {http://arxiv.org/abs/1104.1255} {arXiv:1104.1255 [hep-ph]} \BibitemShut
  {NoStop}%
\bibitem [{\citenamefont {Franzosi}\ and\ \citenamefont
  {Foadi}(2012)}]{Franzosi:2012ih}%
  \BibitemOpen
  \bibfield  {author} {\bibinfo {author} {\bibfnamefont {D.~B.}\ \bibnamefont
  {Franzosi}}\ and\ \bibinfo {author} {\bibfnamefont {R.}~\bibnamefont
  {Foadi}},\ }\href@noop {} {\  (\bibinfo {year} {2012})},\ \Eprint
  {http://arxiv.org/abs/1209.5913} {arXiv:1209.5913 [hep-ph]} \BibitemShut
  {NoStop}%
\bibitem [{\citenamefont {Foadi}\ \emph {et~al.}(2012)\citenamefont {Foadi},
  \citenamefont {Frandsen},\ and\ \citenamefont {Sannino}}]{Foadi:2012bb}%
  \BibitemOpen
  \bibfield  {author} {\bibinfo {author} {\bibfnamefont {R.}~\bibnamefont
  {Foadi}}, \bibinfo {author} {\bibfnamefont {M.~T.}\ \bibnamefont {Frandsen}},
  \ and\ \bibinfo {author} {\bibfnamefont {F.}~\bibnamefont {Sannino}},\ }\href
  {\doibase 10.1103/PhysRevD.87.095001} {\  (\bibinfo {year} {2012}),\
  10.1103/PhysRevD.87.095001},\ \Eprint {http://arxiv.org/abs/1211.1083}
  {arXiv:1211.1083 [hep-ph]} \BibitemShut {NoStop}%
\bibitem [{\citenamefont {Ryttov}\ and\ \citenamefont
  {Sannino}(2008{\natexlab{b}})}]{Ryttov:2007cx}%
  \BibitemOpen
  \bibfield  {author} {\bibinfo {author} {\bibfnamefont {T.~A.}\ \bibnamefont
  {Ryttov}}\ and\ \bibinfo {author} {\bibfnamefont {F.}~\bibnamefont
  {Sannino}},\ }\href {\doibase 10.1103/PhysRevD.78.065001} {\bibfield
  {journal} {\bibinfo  {journal} {Phys.Rev.}\ }\textbf {\bibinfo {volume}
  {D78}},\ \bibinfo {pages} {065001} (\bibinfo {year} {2008}{\natexlab{b}})},\
  \Eprint {http://arxiv.org/abs/0711.3745} {arXiv:0711.3745 [hep-th]}
  \BibitemShut {NoStop}%
\bibitem [{\citenamefont {Pica}\ and\ \citenamefont
  {Sannino}(2011{\natexlab{a}})}]{Pica:2010mt}%
  \BibitemOpen
  \bibfield  {author} {\bibinfo {author} {\bibfnamefont {C.}~\bibnamefont
  {Pica}}\ and\ \bibinfo {author} {\bibfnamefont {F.}~\bibnamefont {Sannino}},\
  }\href {\doibase 10.1103/PhysRevD.83.116001} {\bibfield  {journal} {\bibinfo
  {journal} {Phys.Rev.}\ }\textbf {\bibinfo {volume} {D83}},\ \bibinfo {pages}
  {116001} (\bibinfo {year} {2011}{\natexlab{a}})},\ \Eprint
  {http://arxiv.org/abs/1011.3832} {arXiv:1011.3832 [hep-ph]} \BibitemShut
  {NoStop}%
\bibitem [{\citenamefont {Pica}\ and\ \citenamefont
  {Sannino}(2011{\natexlab{b}})}]{Pica:2010xq}%
  \BibitemOpen
  \bibfield  {author} {\bibinfo {author} {\bibfnamefont {C.}~\bibnamefont
  {Pica}}\ and\ \bibinfo {author} {\bibfnamefont {F.}~\bibnamefont {Sannino}},\
  }\href {\doibase 10.1103/PhysRevD.83.035013} {\bibfield  {journal} {\bibinfo
  {journal} {Phys.Rev.}\ }\textbf {\bibinfo {volume} {D83}},\ \bibinfo {pages}
  {035013} (\bibinfo {year} {2011}{\natexlab{b}})},\ \Eprint
  {http://arxiv.org/abs/1011.5917} {arXiv:1011.5917 [hep-ph]} \BibitemShut
  {NoStop}%
\bibitem [{\citenamefont {Sannino}\ and\ \citenamefont
  {Schechter}(1995)}]{Sannino:1995ik}%
  \BibitemOpen
  \bibfield  {author} {\bibinfo {author} {\bibfnamefont {F.}~\bibnamefont
  {Sannino}}\ and\ \bibinfo {author} {\bibfnamefont {J.}~\bibnamefont
  {Schechter}},\ }\href {\doibase 10.1103/PhysRevD.52.96} {\bibfield  {journal}
  {\bibinfo  {journal} {Phys.Rev.}\ }\textbf {\bibinfo {volume} {D52}},\
  \bibinfo {pages} {96} (\bibinfo {year} {1995})},\ \Eprint
  {http://arxiv.org/abs/hep-ph/9501417} {arXiv:hep-ph/9501417 [hep-ph]}
  \BibitemShut {NoStop}%
\bibitem [{\citenamefont {Harada}\ \emph {et~al.}(1996)\citenamefont {Harada},
  \citenamefont {Sannino},\ and\ \citenamefont {Schechter}}]{Harada:1995dc}%
  \BibitemOpen
  \bibfield  {author} {\bibinfo {author} {\bibfnamefont {M.}~\bibnamefont
  {Harada}}, \bibinfo {author} {\bibfnamefont {F.}~\bibnamefont {Sannino}}, \
  and\ \bibinfo {author} {\bibfnamefont {J.}~\bibnamefont {Schechter}},\ }\href
  {\doibase 10.1103/PhysRevD.54.1991} {\bibfield  {journal} {\bibinfo
  {journal} {Phys.Rev.}\ }\textbf {\bibinfo {volume} {D54}},\ \bibinfo {pages}
  {1991} (\bibinfo {year} {1996})},\ \Eprint
  {http://arxiv.org/abs/hep-ph/9511335} {arXiv:hep-ph/9511335 [hep-ph]}
  \BibitemShut {NoStop}%
\bibitem [{\citenamefont {Harada}\ \emph {et~al.}(1997)\citenamefont {Harada},
  \citenamefont {Sannino},\ and\ \citenamefont {Schechter}}]{Harada:1996wr}%
  \BibitemOpen
  \bibfield  {author} {\bibinfo {author} {\bibfnamefont {M.}~\bibnamefont
  {Harada}}, \bibinfo {author} {\bibfnamefont {F.}~\bibnamefont {Sannino}}, \
  and\ \bibinfo {author} {\bibfnamefont {J.}~\bibnamefont {Schechter}},\ }\href
  {\doibase 10.1103/PhysRevLett.78.1603} {\bibfield  {journal} {\bibinfo
  {journal} {Phys.Rev.Lett.}\ }\textbf {\bibinfo {volume} {78}},\ \bibinfo
  {pages} {1603} (\bibinfo {year} {1997})}\BibitemShut {NoStop}%
\bibitem [{\citenamefont {Black}\ \emph {et~al.}(1999)\citenamefont {Black},
  \citenamefont {Fariborz}, \citenamefont {Sannino},\ and\ \citenamefont
  {Schechter}}]{Black:1998wt}%
  \BibitemOpen
  \bibfield  {author} {\bibinfo {author} {\bibfnamefont {D.}~\bibnamefont
  {Black}}, \bibinfo {author} {\bibfnamefont {A.~H.}\ \bibnamefont {Fariborz}},
  \bibinfo {author} {\bibfnamefont {F.}~\bibnamefont {Sannino}}, \ and\
  \bibinfo {author} {\bibfnamefont {J.}~\bibnamefont {Schechter}},\ }\href
  {\doibase 10.1103/PhysRevD.59.074026} {\bibfield  {journal} {\bibinfo
  {journal} {Phys.Rev.}\ }\textbf {\bibinfo {volume} {D59}},\ \bibinfo {pages}
  {074026} (\bibinfo {year} {1999})},\ \Eprint
  {http://arxiv.org/abs/hep-ph/9808415} {arXiv:hep-ph/9808415 [hep-ph]}
  \BibitemShut {NoStop}%
\bibitem [{\citenamefont {Colangelo}\ \emph {et~al.}(2001)\citenamefont
  {Colangelo}, \citenamefont {Gasser},\ and\ \citenamefont
  {Leutwyler}}]{Colangelo:2001df}%
  \BibitemOpen
  \bibfield  {author} {\bibinfo {author} {\bibfnamefont {G.}~\bibnamefont
  {Colangelo}}, \bibinfo {author} {\bibfnamefont {J.}~\bibnamefont {Gasser}}, \
  and\ \bibinfo {author} {\bibfnamefont {H.}~\bibnamefont {Leutwyler}},\ }\href
  {\doibase 10.1016/S0550-3213(01)00147-X} {\bibfield  {journal} {\bibinfo
  {journal} {Nucl.Phys.}\ }\textbf {\bibinfo {volume} {B603}},\ \bibinfo
  {pages} {125} (\bibinfo {year} {2001})}\BibitemShut {NoStop}%
\bibitem [{\citenamefont {Ananthanarayan}\ \emph {et~al.}(2001)\citenamefont
  {Ananthanarayan}, \citenamefont {Colangelo}, \citenamefont {Gasser},\ and\
  \citenamefont {Leutwyler}}]{Ananthanarayan:2000ht}%
  \BibitemOpen
  \bibfield  {author} {\bibinfo {author} {\bibfnamefont {B.}~\bibnamefont
  {Ananthanarayan}}, \bibinfo {author} {\bibfnamefont {G.}~\bibnamefont
  {Colangelo}}, \bibinfo {author} {\bibfnamefont {J.}~\bibnamefont {Gasser}}, \
  and\ \bibinfo {author} {\bibfnamefont {H.}~\bibnamefont {Leutwyler}},\ }\href
  {\doibase 10.1016/S0370-1573(01)00009-6} {\bibfield  {journal} {\bibinfo
  {journal} {Phys.Rept.}\ }\textbf {\bibinfo {volume} {353}},\ \bibinfo {pages}
  {207} (\bibinfo {year} {2001})},\ \Eprint
  {http://arxiv.org/abs/hep-ph/0005297} {arXiv:hep-ph/0005297 [hep-ph]}
  \BibitemShut {NoStop}%
\bibitem [{\citenamefont {Kaminski}\ \emph {et~al.}(2003)\citenamefont
  {Kaminski}, \citenamefont {Lesniak},\ and\ \citenamefont
  {Loiseau}}]{Kaminski:2002pe}%
  \BibitemOpen
  \bibfield  {author} {\bibinfo {author} {\bibfnamefont {R.}~\bibnamefont
  {Kaminski}}, \bibinfo {author} {\bibfnamefont {L.}~\bibnamefont {Lesniak}}, \
  and\ \bibinfo {author} {\bibfnamefont {B.}~\bibnamefont {Loiseau}},\ }\href
  {\doibase 10.1016/S0370-2693(02)03021-6} {\bibfield  {journal} {\bibinfo
  {journal} {Phys.Lett.}\ }\textbf {\bibinfo {volume} {B551}},\ \bibinfo
  {pages} {241} (\bibinfo {year} {2003})},\ \Eprint
  {http://arxiv.org/abs/hep-ph/0210334} {arXiv:hep-ph/0210334 [hep-ph]}
  \BibitemShut {NoStop}%
\bibitem [{\citenamefont {Caprini}\ \emph {et~al.}(2006)\citenamefont
  {Caprini}, \citenamefont {Colangelo},\ and\ \citenamefont
  {Leutwyler}}]{Caprini:2005zr}%
  \BibitemOpen
  \bibfield  {author} {\bibinfo {author} {\bibfnamefont {I.}~\bibnamefont
  {Caprini}}, \bibinfo {author} {\bibfnamefont {G.}~\bibnamefont {Colangelo}},
  \ and\ \bibinfo {author} {\bibfnamefont {H.}~\bibnamefont {Leutwyler}},\
  }\href {\doibase 10.1103/PhysRevLett.96.132001} {\bibfield  {journal}
  {\bibinfo  {journal} {Phys.Rev.Lett.}\ }\textbf {\bibinfo {volume} {96}},\
  \bibinfo {pages} {132001} (\bibinfo {year} {2006})},\ \Eprint
  {http://arxiv.org/abs/hep-ph/0512364} {arXiv:hep-ph/0512364 [hep-ph]}
  \BibitemShut {NoStop}%
\bibitem [{\citenamefont {Garcia-Martin}\ \emph {et~al.}(2011)\citenamefont
  {Garcia-Martin}, \citenamefont {Kaminski}, \citenamefont {Pelaez},
  \citenamefont {Ruiz~de Elvira},\ and\ \citenamefont
  {Yndurain}}]{GarciaMartin:2011cn}%
  \BibitemOpen
  \bibfield  {author} {\bibinfo {author} {\bibfnamefont {R.}~\bibnamefont
  {Garcia-Martin}}, \bibinfo {author} {\bibfnamefont {R.}~\bibnamefont
  {Kaminski}}, \bibinfo {author} {\bibfnamefont {J.}~\bibnamefont {Pelaez}},
  \bibinfo {author} {\bibfnamefont {J.}~\bibnamefont {Ruiz~de Elvira}}, \ and\
  \bibinfo {author} {\bibfnamefont {F.}~\bibnamefont {Yndurain}},\ }\href
  {\doibase 10.1103/PhysRevD.83.074004} {\bibfield  {journal} {\bibinfo
  {journal} {Phys.Rev.}\ }\textbf {\bibinfo {volume} {D83}},\ \bibinfo {pages}
  {074004} (\bibinfo {year} {2011})},\ \Eprint {http://arxiv.org/abs/1102.2183}
  {arXiv:1102.2183 [hep-ph]} \BibitemShut {NoStop}%
\bibitem [{\citenamefont {Moussallam}(2011)}]{Moussallam:2011zg}%
  \BibitemOpen
  \bibfield  {author} {\bibinfo {author} {\bibfnamefont {B.}~\bibnamefont
  {Moussallam}},\ }\href {\doibase 10.1140/epjc/s10052-011-1814-z} {\bibfield
  {journal} {\bibinfo  {journal} {Eur.Phys.J.}\ }\textbf {\bibinfo {volume}
  {C71}},\ \bibinfo {pages} {1814} (\bibinfo {year} {2011})},\ \Eprint
  {http://arxiv.org/abs/1110.6074} {arXiv:1110.6074 [hep-ph]} \BibitemShut
  {NoStop}%
\bibitem [{\citenamefont {Roy}(1971)}]{Roy:1971tc}%
  \BibitemOpen
  \bibfield  {author} {\bibinfo {author} {\bibfnamefont {S.}~\bibnamefont
  {Roy}},\ }\href {\doibase 10.1016/0370-2693(71)90724-6} {\bibfield  {journal}
  {\bibinfo  {journal} {Phys.Lett.}\ }\textbf {\bibinfo {volume} {B36}},\
  \bibinfo {pages} {353} (\bibinfo {year} {1971})}\BibitemShut {NoStop}%
\bibitem [{\citenamefont {Johnson}\ and\ \citenamefont
  {Teller}(1955)}]{Johnson:1955zz}%
  \BibitemOpen
  \bibfield  {author} {\bibinfo {author} {\bibfnamefont {M.}~\bibnamefont
  {Johnson}}\ and\ \bibinfo {author} {\bibfnamefont {E.}~\bibnamefont
  {Teller}},\ }\href {\doibase 10.1103/PhysRev.98.783} {\bibfield  {journal}
  {\bibinfo  {journal} {Phys.Rev.}\ }\textbf {\bibinfo {volume} {98}},\
  \bibinfo {pages} {783} (\bibinfo {year} {1955})}\BibitemShut {NoStop}%
\bibitem [{\citenamefont {Gell-Mann}\ and\ \citenamefont
  {Levy}(1960)}]{GellMann:1960np}%
  \BibitemOpen
  \bibfield  {author} {\bibinfo {author} {\bibfnamefont {M.}~\bibnamefont
  {Gell-Mann}}\ and\ \bibinfo {author} {\bibfnamefont {M.}~\bibnamefont
  {Levy}},\ }\href {\doibase 10.1007/BF02859738} {\bibfield  {journal}
  {\bibinfo  {journal} {Nuovo Cim.}\ }\textbf {\bibinfo {volume} {16}},\
  \bibinfo {pages} {705} (\bibinfo {year} {1960})}\BibitemShut {NoStop}%
\bibitem [{\citenamefont {Sannino}(2013)}]{Sannino:2013wla}%
  \BibitemOpen
  \bibfield  {author} {\bibinfo {author} {\bibfnamefont {F.}~\bibnamefont
  {Sannino}},\ }\href@noop {} {\  (\bibinfo {year} {2013})},\ \Eprint
  {http://arxiv.org/abs/1306.6346} {arXiv:1306.6346 [hep-ph]} \BibitemShut
  {NoStop}%
\bibitem [{\citenamefont {Degrassi}\ \emph {et~al.}(2012)\citenamefont
  {Degrassi}, \citenamefont {Di~Vita}, \citenamefont {Elias-Miro},
  \citenamefont {Espinosa}, \citenamefont {Giudice} \emph
  {et~al.}}]{Degrassi:2012ry}%
  \BibitemOpen
  \bibfield  {author} {\bibinfo {author} {\bibfnamefont {G.}~\bibnamefont
  {Degrassi}}, \bibinfo {author} {\bibfnamefont {S.}~\bibnamefont {Di~Vita}},
  \bibinfo {author} {\bibfnamefont {J.}~\bibnamefont {Elias-Miro}}, \bibinfo
  {author} {\bibfnamefont {J.~R.}\ \bibnamefont {Espinosa}}, \bibinfo {author}
  {\bibfnamefont {G.~F.}\ \bibnamefont {Giudice}},  \emph {et~al.},\ }\href
  {\doibase 10.1007/JHEP08(2012)098} {\bibfield  {journal} {\bibinfo  {journal}
  {JHEP}\ }\textbf {\bibinfo {volume} {1208}},\ \bibinfo {pages} {098}
  (\bibinfo {year} {2012})},\ \Eprint {http://arxiv.org/abs/1205.6497}
  {arXiv:1205.6497 [hep-ph]} \BibitemShut {NoStop}%
\bibitem [{\citenamefont {Antipin}\ \emph
  {et~al.}(2013{\natexlab{a}})\citenamefont {Antipin}, \citenamefont {Gillioz},
  \citenamefont {Krog}, \citenamefont {M¿lgaard},\ and\ \citenamefont
  {Sannino}}]{Antipin:2013sga}%
  \BibitemOpen
  \bibfield  {author} {\bibinfo {author} {\bibfnamefont {O.}~\bibnamefont
  {Antipin}}, \bibinfo {author} {\bibfnamefont {M.}~\bibnamefont {Gillioz}},
  \bibinfo {author} {\bibfnamefont {J.}~\bibnamefont {Krog}}, \bibinfo {author}
  {\bibfnamefont {E.}~\bibnamefont {M¿lgaard}}, \ and\ \bibinfo {author}
  {\bibfnamefont {F.}~\bibnamefont {Sannino}},\ }\href@noop {} {\  (\bibinfo
  {year} {2013}{\natexlab{a}})},\ \Eprint {http://arxiv.org/abs/1306.3234}
  {arXiv:1306.3234 [hep-ph]} \BibitemShut {NoStop}%
\bibitem [{\citenamefont {Antipin}\ \emph
  {et~al.}(2013{\natexlab{b}})\citenamefont {Antipin}, \citenamefont {Gillioz},
  \citenamefont {Mølgaard},\ and\ \citenamefont {Sannino}}]{Antipin:2013pya}%
  \BibitemOpen
  \bibfield  {author} {\bibinfo {author} {\bibfnamefont {O.}~\bibnamefont
  {Antipin}}, \bibinfo {author} {\bibfnamefont {M.}~\bibnamefont {Gillioz}},
  \bibinfo {author} {\bibfnamefont {E.}~\bibnamefont {Mølgaard}}, \ and\
  \bibinfo {author} {\bibfnamefont {F.}~\bibnamefont {Sannino}},\ }\href
  {\doibase 10.1103/PhysRevD.87.125017} {\  (\bibinfo {year}
  {2013}{\natexlab{b}}),\ 10.1103/PhysRevD.87.125017},\ \Eprint
  {http://arxiv.org/abs/1303.1525} {arXiv:1303.1525 [hep-th]} \BibitemShut
  {NoStop}%
\bibitem [{\citenamefont {Weinberg}(1979)}]{Weinberg:1979bn}%
  \BibitemOpen
  \bibfield  {author} {\bibinfo {author} {\bibfnamefont {S.}~\bibnamefont
  {Weinberg}},\ }\href {\doibase 10.1103/PhysRevD.19.1277} {\bibfield
  {journal} {\bibinfo  {journal} {Phys.Rev.}\ }\textbf {\bibinfo {volume}
  {D19}},\ \bibinfo {pages} {1277} (\bibinfo {year} {1979})}\BibitemShut
  {NoStop}%
\bibitem [{\citenamefont {Susskind}(1979)}]{Susskind:1978ms}%
  \BibitemOpen
  \bibfield  {author} {\bibinfo {author} {\bibfnamefont {L.}~\bibnamefont
  {Susskind}},\ }\href {\doibase 10.1103/PhysRevD.20.2619} {\bibfield
  {journal} {\bibinfo  {journal} {Phys.Rev.}\ }\textbf {\bibinfo {volume}
  {D20}},\ \bibinfo {pages} {2619} (\bibinfo {year} {1979})}\BibitemShut
  {NoStop}%
\bibitem [{\citenamefont {Belyaev}\ \emph {et~al.}(2013)\citenamefont
  {Belyaev}, \citenamefont {Brown}, \citenamefont {Foadi},\ and\ \citenamefont
  {Frandsen}}]{Belyaev:2013ida}%
  \BibitemOpen
  \bibfield  {author} {\bibinfo {author} {\bibfnamefont {A.}~\bibnamefont
  {Belyaev}}, \bibinfo {author} {\bibfnamefont {M.~S.}\ \bibnamefont {Brown}},
  \bibinfo {author} {\bibfnamefont {R.}~\bibnamefont {Foadi}}, \ and\ \bibinfo
  {author} {\bibfnamefont {M.~T.}\ \bibnamefont {Frandsen}},\ }\href@noop {} {\
   (\bibinfo {year} {2013})},\ \Eprint {http://arxiv.org/abs/1309.2097}
  {arXiv:1309.2097 [hep-ph]} \BibitemShut {NoStop}%
\end{thebibliography}%

\end{document}